 \documentclass[10pt,pra,aps,a4paper,oneside,twocolumn,superscriptaddress,showpacs]{revtex4-1}

\usepackage[english]{babel}
\usepackage[latin1]{inputenc}
\usepackage[T1]{fontenc}
\usepackage{lmodern}
\usepackage{ulem}
\usepackage{dcolumn}
\usepackage{subfigure}
\usepackage{footnote}
\usepackage{multirow}
\usepackage{amsmath,amsfonts,amssymb}

\usepackage{graphicx}  % standard LaTeX graphics tool
\usepackage{latexsym}   % useful for coding complex math
\usepackage{color}

\begin{document}

\title{Fano resonances and   fluorescence enhancement of a dipole emitter   near a plasmonic nanoshell}

\author{\firstname{Tiago}  J. \surname{Arruda}}
\email{tiagojarruda@gmail.com}
\affiliation{Instituto de F\'isica de S\~ao Carlos,
Universidade de S\~ao Paulo, 13566-590 S\~ao Carlos, S\~ao Paulo, Brazil}

\author{\firstname{Romain}  \surname{Bachelard}}

\affiliation{Departamento de F\'isica,
Universidade Federal de S\~ao Carlos, 13565-905 S\~ao Carlos, S\~ao Paulo, Brazil}

\author{\firstname{John}  \surname{Weiner}}
\affiliation{Instituto de F\'isica de S\~ao Carlos,
Universidade de S\~ao Paulo, 13566-590 S\~ao Carlos, S\~ao Paulo, Brazil}

\author{\firstname{Sebastian}  \surname{Slama}}
\affiliation{Physikalisches Institut, 
Eberhardt-Karls-Universit\"{a}t T\"{u}bingen, D-72076 T\"{u}bingen, Germany}

\author{\firstname{Philippe}  W. \surname{Courteille}}

\affiliation{Instituto de F\'isica de S\~ao Carlos,
Universidade de S\~ao Paulo, 13566-590 S\~ao Carlos, S\~ao Paulo, Brazil}

\begin{abstract}

We analytically study the spontaneous emission of a single optical dipole emitter in the vicinity of a plasmonic nanoshell, based on the Lorenz-Mie theory. 
We show that the fluorescence enhancement due to the coupling between optical emitter and sphere can be tuned by the   thickness ratio   of the core-shell nanosphere and  by the distance between the quantum emitter and its surface. 
In particular, we demonstrate that both the enhancement and quenching of the fluorescence intensity are associated with plasmonic Fano resonances induced by near- and far-field interactions.
These Fano resonances have asymmetry parameters whose signs depend on the orientation of the dipole with respect to the spherical nanoshell.
We also show that if the atomic dipole is oriented tangentially to the nanoshell, the interaction exhibits saddle points in the near-field energy flow.  
This results in a Lorentzian fluorescence enhancement response in the near field and a Fano lineshape in the far field. 
The signatures of this interaction may have interesting applications for sensing the presence and the orientation of optical emitters in close proximity to plasmonic nanoshells.  

\end{abstract}


\maketitle

%\tableofcontents

\section{Introduction}

The Fano resonance is one of the hallmarks of interference between discrete and continuous states in open quantum systems.
It was originally conceived as an interference between a transition to a bound state, coupled weakly to a continuum, and a transition directly to the same continuum~\cite{Fano_PhysRev124_1961}.
Being a wave phenomenon, the Fano effect can also be understood as weak coupling between two classical oscillators driven by an external harmonic force~\cite{Nussenzveig_AmJPhys70_2002,Kivshar_RevModPhys82_2010}.
In plasmonics, it arises from the interference between a localized, narrow subradiant (dark) mode and a spectrally broad superradiant (bright) mode acting as a background~\cite{Kivshar_RevModPhys82_2010}.
Recently, with the advent of nanoplasmonics and metamaterials, the Fano interference has become an essential tool for tailoring and controlling light-matter interaction at the nanoscale~\cite{Luk_NatMat9_2010}, such as the plasmonic cloaking technique~\cite{Limonov_SciRep5_2015}, comblike scattering response~\cite{Monticone_PhysRevLett110_2013,Arruda_PhysRevA92_2015}, off-resonance field enhancement~\cite{Miroshnichenko_PhysRevA81_2010,Arruda_PhysRevA87_2013}, optical vortices~\cite{Miroshnichenko_OptLett37_2012}, superscattering~\cite{Kivshar_OptLett38_2013}, and atom-plasmon coupling~\cite{Slama_NatPhys10_2014}.

In plasmonic mesoscopic systems, the Fano effect can appear due to the interaction of a localized plasmon resonance with a broad Mie scattering resonance~\cite{Kivshar_RevModPhys82_2010}.
Within the Lorenz-Mie scattering theory, the Fano effect can be observed in (i) the interference between multipoles of different orders (e.g., dipole-quadrupole interference)~\cite{Luk_NatMat9_2010} or (ii) multipoles of the same orders (e.g., dipole-dipole interference), which is sometimes referred to as unconventional Fano resonance~\cite{Arruda_PhysRevA87_2013}.
The former is easily obtained for specific directions of scattering and generally does not depend on the material properties of the scatterer~\cite{Kivshar_JOpt15_2013}; the latter, however, is less general and can be achieved only for specific geometries, such as layered~\cite{Arruda_PhysRevA87_2013,Monticone_PhysRevLett110_2013} or high permittivity particles~\cite{Tribelsky_EurophyLett97_2012,Limonov_OptExpress21_2013,Tribelsky_PhysRevA93_2016}, and it is usually independent of the scattering direction.
Interestingly enough, the Fano effect in Lorenz-Mie scattering can be associated with the formation of optical vortices and saddle points in the energy flow around the particles~\cite{Kivshar_JOpt15_2013}.

Here, we study the impact of a Fano resonance of a plasmonic nanoshell on an optical dipole emitter in its vicinity.  
The presence of a nanostructure is known to enhance the spontaneous-emission rate of atoms~\cite{Kerker_AppOpt19_1980,Chew_JCPhys87_1987,Klimov_OptComm211_2002,Vidal_PhysRevLett112_2014,Zadkov_PhysRevA90_2014}, and many approaches have been developed to maximize~\cite{Datsyuk_PhysRevA75_2007,Szilar_PhysRevB94_2016} or minimize~\cite{Farina_PhysRevA87_2013,Mahdifar_PhysRevA94_2016} the coupling between the emitter and surface electromagnetic modes, using, e.g., engineered hyperbolic metamaterials~\cite{Liu_NatNano9_2014}.
  Furthermore, there have been various experimental and theoretical studies that have pointed out the appearance of the Fano effect due to the overlap of a plasmon resonance with the Lorentzian response of quantum emitters~\cite{Slama_NatPhys10_2014,Simovski_Photonics2_2015}. 
Here we address the impact of a plasmonic Fano resonance, arising from Lorenz-Mie scattering, in the vicinity of nanoshells.
 
With this aim, we explore the unconventional Fano effect in the Lorenz-Mie scattering by considering a core-shell nanoparticle with realistic optical parameters.
In order to calculate the spontaneous-emission rate of an optical emitter near the sphere, we apply a technique that is well established in classical electrodynamics, based on multipole expansion of classical electromagnetic fields in terms of vector spherical wave functions~\cite{Chew_JCPhys87_1987,Ruppin_JCPhys76_1982,Letokhov_JModOpt43_2_1996,Welsch_PhysRevA64_2001}. 
Using the full-wave Lorenz-Mie theory~\cite{Bohren_Book_1983}, we show that the plasmonic Fano resonance of a silver nanoshell can lead to a large enhancement in the radiative spontaneous-emission rate of an atomic dipole.  
We show that the maximum fluorescence enhancement occurs for a certain distance between the optical emitter and the nanoshell, and it is controlled by near-field interactions.
This distance is approximately the same irrespective of the dipole orientation, and it depends on ohmic losses within the nanoshell.  
The strong fluorescence enhancement response is achieved when the dipole is oriented normal to the spherical nanoshell surface.
We also verify that the fluorescence enhancement of the dipole orientated tangentially to the spherical nanoshell is sensitive to optical vortices and saddle points in the near-field.
As the distance to the spherical surface increases, the fluorescence enhancement as a function of wavelength changes from a  symmetric Lorentzian line shape in the near field to an asymmetric Fano resonance when passing through a saddle point in the energy flow.

This paper is organized as follows.
In Sec.~\ref{Theory}, we present the theory of an atomic dipole in the vicinity of an arbitrary sphere in the framework of the Lorenz-Mie theory.
The analytical expressions related to this section are provided in Appendices~\ref{Lorenz-Mie}, \ref{Gamma-calculation}, and \ref{Intensity}.
In Sec.~\ref{Numerical}, we numerically calculate the spontaneous emission rates and the fluorescence enhancement associated with an atomic dipole near a plasmonic silver nanoshell.
We discuss our main results and conclude in Sec.~\ref{Conclusion}.

\section{Decay rates and fluorescence enhancement in the vicinity of a sphere}
\label{Theory}
  
In quantum electrodynamics, the first-order perturbation theory is the standard approach to calculate the variation on linewidth and energy-level shift of a single atom due to the environment~\cite{Wylie_PhysRevA30_1984}, which is generally referred to as the Purcell effect~\cite{Wildea_SurfSciRep70_2015,Kivshar_SciRep5_2015}.
In the weak-coupling regime, the spontaneous emission of an atom follows Fermi's golden rule, in which the atom decays exponentially to its ground state.
A remarkable feature of this approximation is that the emission rate of an atom in the vicinity of a body, normalized by the spontaneous emission rate in vacuum, $\Gamma/\Gamma_0$, can be calculated in the framework of classical electrodynamics~\cite{Letokhov_JModOpt43_2_1996,Wildea_SurfSciRep70_2015}. 
Indeed, by taking the total radiated power by a classical dipole at frequency $\omega_0$ in the presence of a body, normalized by the corresponding radiated power in vacuum, $P/P_0$, one can formally demonstrate that $P/P_0=\Gamma/\Gamma_0$, where $\Gamma/\Gamma_0$ is calculated at the transition frequency $\omega_0$~\cite{Chew_JCPhys87_1987,Datsyuk_PhysRevA75_2007}.

\begin{figure}[htbp]
\centerline{\includegraphics[width=1.1\columnwidth]{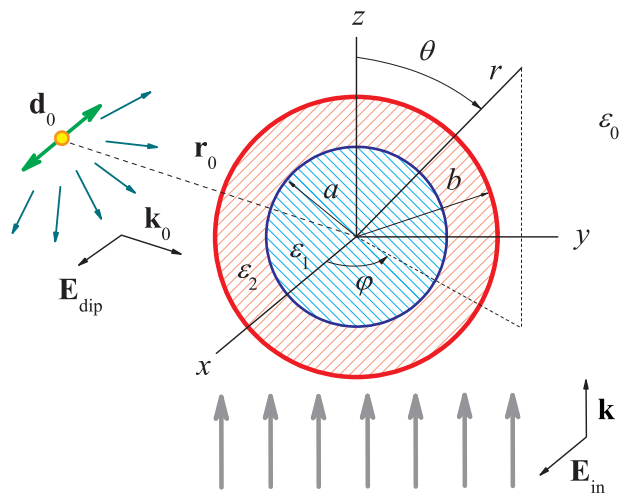}}
\caption{An electric dipole emitter in the vicinity of an illuminated coated sphere with inner radius $a$ and outer radius $b$.
There are two basic orientations for the electric dipole moment at $\mathbf{r}_0$: orthogonal ($\mathbf{d}_{0}^{\perp})$ and parallel ($\mathbf{d}_{0}^{||}$) to the spherical surface.
The sphere has optical properties $(\varepsilon_1,\mu_1)$ for the core $(0<r\leq a)$ and $(\varepsilon_2,\mu_2)$ for the shell $(a<r\leq b)$, where $\varepsilon$ ($\mu$) is the permittivity (permeability).
The surrounding medium is ($\varepsilon_0,\mu_0)$.
The sphere and the dipole emitter are exposed to an incoming electromagnetic wave with electric field $\mathbf{E}_{\rm in}=E_0e^{\imath kz}\hat{\mathbf{x}}$ ($k=\omega\sqrt{\varepsilon_0\mu_0}$). 
The dipole response is represented by the electric field $\mathbf{E}_{\rm dip}$.}\label{fig1}
\end{figure}

Bearing this equivalence in mind~\cite{Aguanno_PhysRevE69_2004}, let us consider an arbitrary coated sphere of inner radius $a$ and outer radius $b$ embedded in a non-dispersive and non-absorbing medium with permittivity $\varepsilon_0$ and permeability $\mu_0$.
The sphere has optical properties $(\varepsilon_1,\mu_1)$ for the core $(r\leq a)$ and $(\varepsilon_2,\mu_2)$ for the shell $(a\leq r\leq b)$, as depicted in Fig.~\ref{fig1}.
Both core and shell consist of isotropic and linear materials, and may have absorption and dispersion that  satisfy the Kramers-Kronig relations~\cite{Welsch_PhysRevA64_2001,Farina_PhysRevA87_2013}.

As illustrated in Fig.~\ref{fig1}, the sphere interacts with a single electric dipole located at position $r_0=|\mathbf{r}_0|>b$ (the coordinate system has its origin at the center of the sphere) with dipole moment $\mathbf{d}_0$ emitting at the fluorescence frequency $\omega_0$. 
There are two processes depicted in Fig.~\ref{fig1}: (i) the spontaneous emission of a dipole emitter near a plasmonic nanoshell; (ii) the excitation of the whole system by an incoming electromagnetic wave from below. 
With respect to the nanosphere, there are two basic orientations for the optical emitter: the dipole moment is orthogonally $(\mathbf{d}_0^{\perp})$ or tangentially $(\mathbf{d}_0^{||})$ oriented in relation to the spherical surface.
The dipole electric field $\mathbf{E}_{\rm dip}^{\mathbf{d}_0}(\mathbf{r},\omega_0)$ impinges on the spherical shell producing the scattered field $\mathbf{E}_{\rm sca}^{\mathbf{d}_0}(\mathbf{r},\omega_0)$ for $r>b$. 
The radiative decay rate of an atomic dipole at the position $\mathbf{r}_0$ can be readily calculated via the normalized radiated power in the surrounding medium $(\varepsilon_0,\mu_0)$, in the presence and absence of the sphere.
The total radiated power is calculated by integrating the radial component of the Poynting vector at the far field $(r\to\infty)$: $P=r^2\int {\rm d}\Omega\mathbf{S}\cdot\hat{\mathbf{r}}\propto r^2\int {\rm d}\Omega|\mathbf{E}_{\rm dip}^{\mathbf{d}_0}+\mathbf{E}_{\rm sca}^{\mathbf{d}_0}|^2$ (for details, see, {e.g.}, Ref.~\cite{Chew_JCPhys87_1987}). 
Using the Green's tensor formalism and the notation of Ref.~\cite{Letokhov_JModOpt43_2_1996}, in both classical and quantum electrodynamics~\cite{Chew_JCPhys87_1987}, the solution for the total decay rate associated with an electric dipole moment ${\mathbf{d}}_0$ can be expressed as 
\begin{align}
\frac{\Gamma_{\mathbf{d}_0}(\omega_0)}{\Gamma_{0}}&=1 + \frac{3}{2k_0^3d_0^2}{\rm Im}\left[\mathbf{d}_0\cdot\overleftrightarrow{\mathbf{G}}_{\rm E}^{\rm sca}(\mathbf{r},\mathbf{r}_0,\omega_0)\cdot\mathbf{d}_0\right]\nonumber\\
&=1+\frac{3}{2k_0^3d_0^2}{\rm Im}\left[\mathbf{d}_0\cdot\mathbf{E}_{\rm sca}^{\mathbf{d}_0}(\mathbf{r},\omega_0)\right],\label{equiv1}
\end{align}
where the scattered electric field, expressed as the electric Green's tensor ``dotted'' into the electric dipole moment, contains the information of the environment (boundary conditions) in which the optical emitter is embedded and $k_0=\omega_0\sqrt{\varepsilon_0\mu_0}$.
Equation~(\ref{equiv1}) takes into account both the radiative ($\Gamma_{\mathbf{d}_0}^{\rm rad}$) and nonradiative ($\Gamma_{\mathbf{d}_0}^{\rm nrad}$) decay rates, which are associated with near-field and far-field interactions, respectively, with a dispersive sphere material.
 
From a quantum perspective, one can identify from Eq.~(\ref{equiv1}) an expression for the projected~\cite{Wildea_SurfSciRep70_2015} (in $\mathbf{d}_0$) electric local density of states (LDOS): $\rho_{\mathbf{d}_0}(\omega_0)\propto{\rm Im}[\mathbf{d}_0\cdot\overleftrightarrow{\mathbf{G}}_{\rm E}(\mathbf{r},\mathbf{r}_0,\omega_0)\cdot\mathbf{d}_0]$, where $\overleftrightarrow{\mathbf{G}}_{\rm E}(\mathbf{r},\mathbf{r}_0,\omega_0)\cdot\mathbf{d}_0=\mathbf{E}_{\rm dip}^{\mathbf{d}_0}(\mathbf{r},\omega_0)+\mathbf{E}_{\rm sca}^{\mathbf{d}_0}(\mathbf{r},\omega_0)$ is the total electric Green's tensor   projected in $\mathbf{d}_0$.  
The full electric LDOS takes into account the three possible directions of the dipole moment $\mathbf{d}_0$ and is defined as the trace of the total electric Green's tensor: $\rho(\omega_0)\propto{\rm Im}{\rm Tr}[\overleftrightarrow{\mathbf{G}}_{\rm E}(\mathbf{r},\mathbf{r}_0,\omega_0)]$.
Calling $\rho_0$ the electric LDOS associated with the emitter in vacuum, one has $\rho(\omega_0)/\rho_0=\Gamma(\omega_0)/{\Gamma_0}$.
It is worth mentioning that in the classical picture the operators must be replaced with the angle average of the corresponding vector functions and the trace operation is simply the spatial mean~\cite{Chew_JCPhys87_1987,Wildea_SurfSciRep70_2015}.	

This theory provides a fully classical computational method to derive a quantum property of a system, {i.e.}, $\Gamma/\Gamma_0$~\cite{Aguanno_PhysRevE69_2004}.
To study the fluorescence enhancement within this theory, we consider that the absorption coefficient of light by the optical emitter depends only on the excitation wavelength.
There are some analytical approaches that take into account the Lorentzian fluorescence spectra of the atomic dipole as a function of the detuning frequency~\cite{Grimm_AdAtomPhys42_2000,Zadkov_PhysRevA85_2012}, which we are not considering here. 
Furthermore, to discuss the fluorescence enhancement factor due the coupling between dipole and  nanoshell, one must assume that the atomic transition does not saturate, such that the dipole is linear, $\mathbf{d}_0\propto\mathbf{E}_{\rm in}(\omega)$~\cite{Dujardin_OptExp16_2008}.  
Notwithstanding these considerations, this analytical theory is in good agreement with experimental data for quantum emitters near dielectric and plasmonic spherical particles~\cite{Gaponenko_JPhysChem116_2012,Bach_ACSNano7_2013}.
For the sake of completeness, the well-known analytical expressions of the electromagnetic fields in the Lorenz-Mie theory, the spontaneous-emission rate, and the intensity enhancement factor are determined in Appendices~\ref{Lorenz-Mie}, \ref{Gamma-calculation} and \ref{Intensity}, respectively.

The fluorescence enhancement factor $F_{\mathbf{d}_0}$ is defined as the ratio between the observed emission intensities in the presence and in the absence of the sphere in the vicinity of the optical emitter~\cite{Dujardin_OptExp16_2008,Gaponenko_JPhysChem116_2012,Bach_ACSNano7_2013}.
In the absence of the sphere, the observed emission intensity is $I_0=|\mathbf{d}_0\cdot\mathbf{E}_{\rm in}|^2\zeta_A Q_0$, where $\zeta_A$ and $Q_0$ are the absorption coefficient (which depends on the emitter polarizability) and the quantum yield of the optical emitter, respectively.
This latter quantity $Q_0$ is related to the ratio between radiative and nonradiative decay rates.
In vacuum, one has $Q_0=1$.
When the sphere is considered in the environment, both the quantum yield and the excitation intensity are modified, so that the observed emission intensity is $I_F=|\mathbf{d}_0\cdot(\mathbf{E}_{\rm in}+\mathbf{E}_{\rm sca})|^2 \zeta_A Q_{\mathbf{d}_0}$. 
Here, we assume that the absorption coefficient of light $\zeta_A$ by an optical dipole emitter is intensity independent and remains the same as in vacuum~\cite{Gaponenko_JPhysChem116_2012}.
We also assume that the quantum yield $Q_{\mathbf{d}_0}$ depends only on the sphere.
Being $\Gamma_{\mathbf{d}_0}=\Gamma_{\mathbf{d}_0}^{\rm rad}+\Gamma_{\mathbf{d}_0}^{\rm nrad}$ and $F_{\mathbf{d}_0}\equiv I_F/I_0$, one has 
\begin{align}
F_{\mathbf{d}_0}^{\omega_0}(r_0,\omega)=\mathcal{G}_{\mathbf{d}_0}(r_0,\omega) Q_{\mathbf{d}_0}^{\omega_0}(r_0),\label{Factor}
\end{align}
where the quantities
\begin{align}
\mathcal{G}_{\mathbf{d}_0}(r_0,\omega)&=\frac{\left\langle \left|\mathbf{d}_0\cdot\left[\mathbf{E}_{\rm in}(\mathbf{r}_0,\omega)+\mathbf{E}_{\rm sca}(\mathbf{r}_0,\omega)\right]\right|^2\right\rangle}{\left\langle\left|\mathbf{d}_0\cdot\mathbf{E}_{\rm in}(\mathbf{r}_0,\omega)\right|^2\right\rangle},\label{G-factor}\\
Q_{\mathbf{d}_0}^{\omega_0}(r_0)&=\frac{\Gamma_{\mathbf{d}_0}^{\rm rad}(r_0,\omega_0)}{\Gamma_{\mathbf{d}_0}^{\rm rad}(r_0,\omega_0)+\Gamma_{\mathbf{d}_0}^{\rm nrad}(r_0,\omega_0)}
\end{align}
are the   averaged   intensity enhancement factor (which depends on $\omega$, the {\it excitation} frequency) and the quantum efficiency (which depends on $\omega_0$, the {\it transition} frequency), respectively, with $\mathbf{E}_{\rm sca}(\omega)$ being the scattered field.
The operator $\langle \cdots \rangle = (1/{4\pi})\int_{-1}^{1}{\rm d}(\cos\theta)\int_0^{2\pi}{\rm d}\varphi (\cdots)$ is the angle average over 4$\pi$ and is analytically calculated in Appendix~\ref{Intensity}.
The quantum yield $Q_{\mathbf{d}_0}^{\omega_0}(r_0)$ can be readily obtained from Eqs.~(\ref{Gamma-perprad})--(\ref{Gamma-paran}) of Appendix~\ref{Gamma-calculation}.
We emphasize that we are considering the weak-coupling regime, so that $Q_{\mathbf{d}_0}^{\omega_0}(r_0)$ is only modified by the presence of the sphere and not by the excitation intensity~\cite{Kivshar_SciRep5_2015}.
The interaction between the atomic dipole and the incoming (scattered) electric fields is encoded in the intensity enhancement factor $\mathcal{G}_{\mathbf{d}_0}(r_0,\omega)$. 
Of course, if there is no sphere interacting with the dipole, then $F_{\mathbf{d}_0}^{\omega_0}(r_0,\omega)=1$, provided that $\mathbf{d}_0$ is parallel to $\mathbf{E}_{\rm in}$.

The theory presented above is general and can be applied to arbitrary spheres and dipoles (quantum dots, atoms, or molecules) below the saturated intensity regime~\cite{Dujardin_OptExp16_2008,Gaponenko_JPhysChem116_2012,Bach_ACSNano7_2013}.
Now, we consider a realistic system for a quantum emitter in the vicinity of a plasmonic nanosphere.
Here, we are interested in a configuration where the presence of a dielectric core strongly modifies the plasmonic scattering response~\cite{Arruda_JOpSocAmA31_2014,Arruda_PhysRevA94_2016,Li_IEEE23_2017,Li_Nanoscale9_2017} and may induce plasmon hybridization~\cite{Nordlander_Science17_2003} and, ultimately, Fano resonances~\cite{Halas_NanoLett10_2010}.
For subwavelength structures, the plasmon hybridization corresponds to an interference between dipole resonances excited, e.g., at the plasmonic shell/surrounding medium interface (broad mode) and at the dielectric core/plasmonic shell interface (narrow mode).
The constructive interference between these two surface modes produces a ``bonding'' mode, whereas the destructive interference between them produces an ``anti-bonding'' mode~\cite{Halas_NanoLett10_2010}.

\section{Atomic dipole in the vicinity of a silver nanoshell}
\label{Numerical}

Let us consider a nanoparticle consisting of a lossless dielectric core with refractive index $n_1=3.5$ and radius $a=50$~nm coated with a dispersive silver (Ag) nanoshell with radius $b=70$~nm.
The Ag dielectric permittivity function is calculated using experimental data and dispersion relations from Refs.~\cite{Christy_PhysRevB6_1972,Shalaev_OptExp16_2008}.
From the standard Lorenz-Mie theory, the far-field extinction, scattering and absorption cross sections associated with a coated sphere are, respectively~\cite{Bohren_Book_1983},
\begin{align}
\sigma_{\rm ext} &= \frac{2\pi}{k^2}\sum_{\ell=1}^{\infty}(2\ell+1){\rm Re}\left[a_{\ell}(\omega)+b_{\ell}(\omega)\right],\label{Qext}\\
\sigma_{\rm sca} &= \frac{2\pi}{k^2}\sum_{\ell=1}^{\infty}(2\ell+1)\left[|a_{\ell}(\omega)|^2+|b_{\ell}(\omega)|^2\right],\\
\sigma_{\rm abs} &= \sigma_{\rm ext}-\sigma_{\rm sca},\label{Qabs}
\end{align}
where $a_{\ell}(\omega)$ and $b_{\ell}(\omega)$ carry the dependence on the geometrical and material parameters of the scatterer (Appendix~\ref{Lorenz-Mie}) and $k=\omega\sqrt{\varepsilon_0\mu_0}$ is the wave number in the surrounding medium.

\begin{figure}[htbp!]
\centerline{\includegraphics[width=\columnwidth]{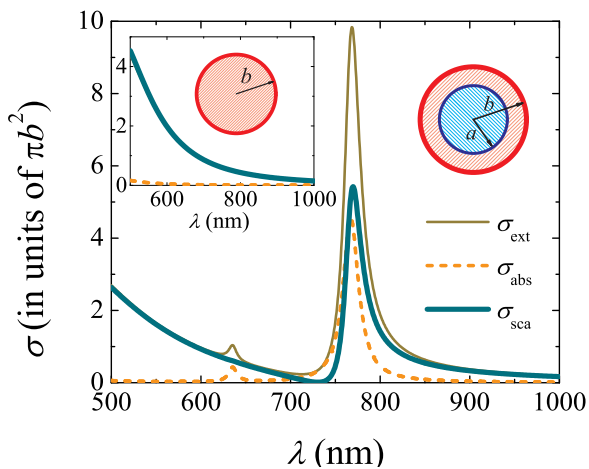}}
\caption{Optical cross sections associated with a silver (Ag) nanoshell illuminated by planes waves as a function of the excitation wavelength.
The lossless dielectric core has refractive index $n_1=3.5$ and radius $a=50$~nm, whereas the Ag shell has radius $b=70$~nm and dispersive permittivity provided by Ref.~\cite{Christy_PhysRevB6_1972}.
The resonant peak in the cross sections around 770~nm is due to plasmon hybridization between the core-shell ($r=a$) and shell-external medium ($r=b$) interface plasmon modes, and have a Lorentzian lineshape for the absorption cross section $\sigma_{\rm abs}$ and a Fano lineshape for the scattering cross section $\sigma_{\rm sca}$.
The inset shows that these peaks vanish for a homogenous Ag nanoparticle with radius $b=70$~nm.
The quadrupolar resonance peak around 635~nm is also due to plasmon hybridization. }\label{fig2}
\end{figure}

In Fig.~\ref{fig2}, we plot the Lorenz-Mie cross sections (in units of $\pi b^2$) in the optical frequency range, so that $0.44<kb<0.88$.
Within this parameter range, the dipole approximation can still be applied to understand the scattering resonances.
In particular, note that the presence of a dielectric core leads to a strong absorption and scattering resonance peaks around 770~nm (compare it with the inset in Fig.~\ref{fig2}).
This resonance, which is due to the core-shell geometry, can be explained by the plasmon hybridization between plasmon modes in the inner and outer Ag spherical surfaces~\cite{Nordlander_Science17_2003}.
Indeed, we choose the geometrical parameters   ($b=70$~nm and thickness ratio $a/b=5/7$)   to obtain a resonance peak within the Ag nanoshell around  $\lambda_0=780$~nm.
The interference between the sphere and cavity plasmon modes gives rise to a Fano line-shape response in the scattering cross section and a Lorentzian (Breit-Wigner) line-shape response in the absorption cross section~\cite{Arruda_PhysRevA87_2013}.  
It is worth mentioning that one could also achieve Fano resonances in the total scattering cross section of a homogeneous high-permittivity dielectric sphere~\cite{Limonov_OptExpress21_2013,Tribelsky_PhysRevA93_2016}, which is not the focus of our study.
 
The scattering profile showed in Fig.~\ref{fig2} can be mainly explained using the Lorenz-Mie coefficient $a_1$ (the electric dipole amplitude): $\sigma_{\rm sca}\approx 6\pi|a_1|^2/k^2$~\cite{Arruda_PhysRevA87_2013,Arruda_PhysRevA92_2015}.
The asymmetric lineshape is due to the constructive and destructive interferences between in and out-of-phase electric fields associated with the surface plasmon waves at the two metal/dielectric surfaces.
This leads to a resonance at $\lambda=770$~nm  and an antiresonance at $\lambda=740$~nm, respectively.
In particular, the antiresonance scattering response due to the presence of the plasmonic nanoshell can be used as a technique to achieve near-invisibility of dielectric nanoparticles within a narrow band of wavelengths~\cite{Farina_PhysRevA87_2013,Limonov_SciRep5_2015}. 

\begin{figure}[htbp!]
\includegraphics[width=\columnwidth]{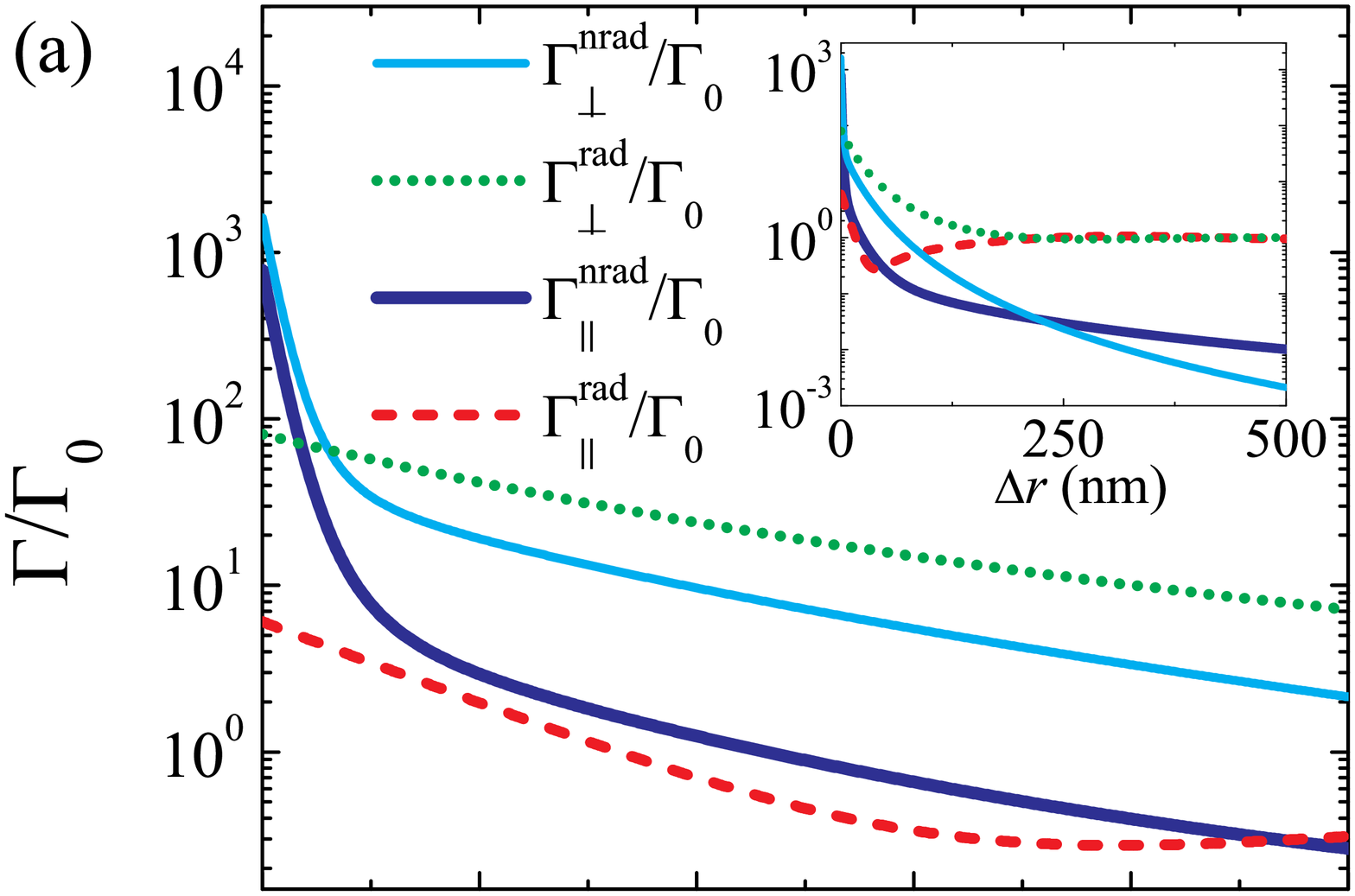}\vspace{-1.8cm}
\includegraphics[width=\columnwidth]{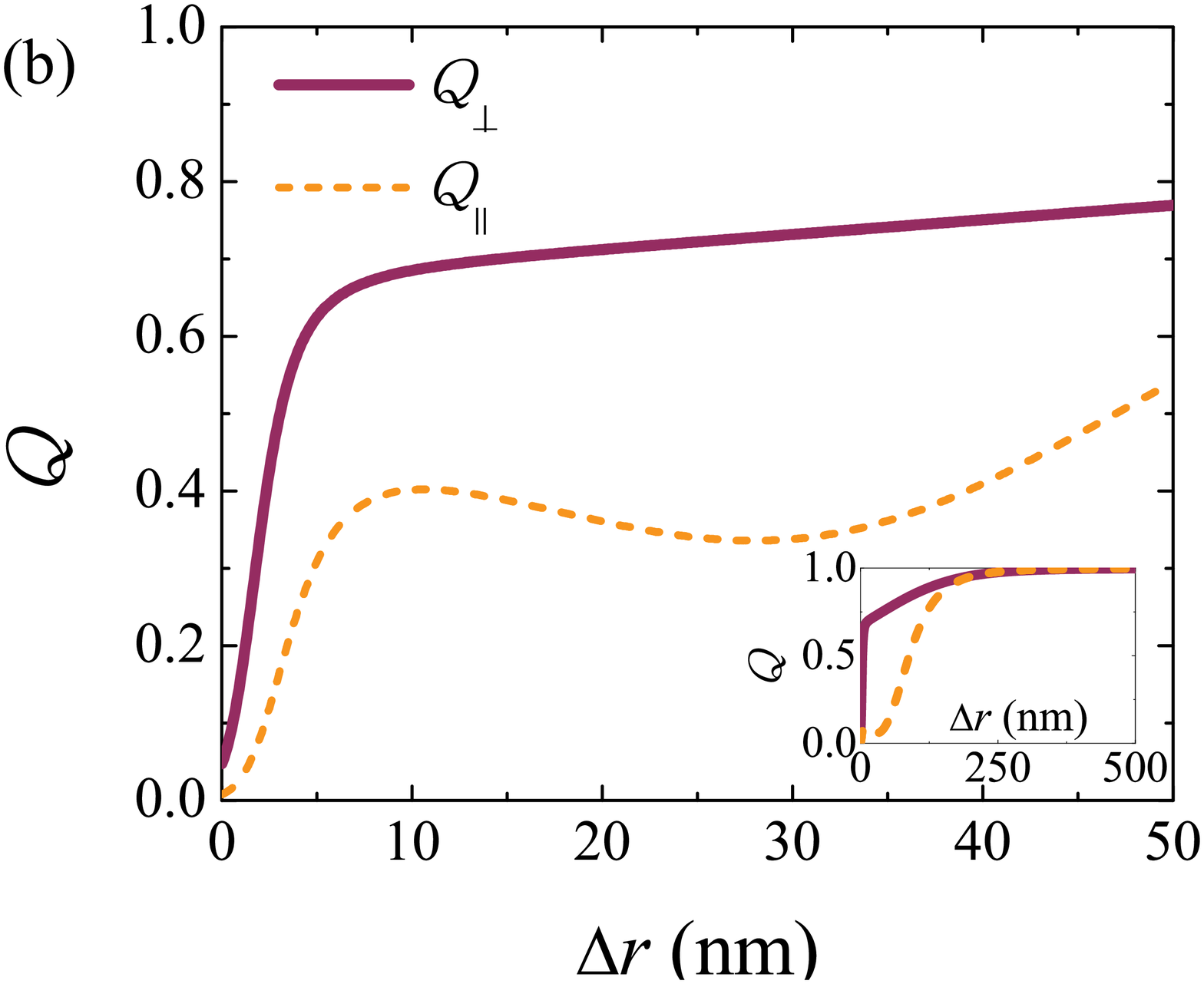}
\caption{Spontaneous-decay rates for an optical emitter with transition wavelength $\lambda_0=780$~nm in the vicinity of a silver nanoshell as a function of the distance to the spherical surface $\Delta r=r_0-b$.
The nanoparticle consists of a dielectric core with $n_1=3.5$ and radius $a=50$~nm and a silver shell with radius $b=70$~nm.
(a) The radiative ($\Gamma^{\rm rad}$) and non-radiative $(\Gamma^{\rm nrad})$ decay rates for a dipole oriented tangentially $(||)$ or normally $(\perp)$ to the spherical surface normalized by the decay rate in vacuum $(\Gamma_0)$.
The inset shows that $\Gamma_{\perp(||)}^{\rm rad}\to \Gamma_0$ and $\Gamma_{\perp(||)}^{\rm nrad}\to 0$  as $\Delta r\gg b$
(b) The corresponding quantum efficiency $Q=\Gamma^{\rm rad}/\Gamma$ for each orientation.
The inset shows that $Q_{\perp(||)}\to 1$ as $\Delta r\gg b$.}\label{fig3}
\end{figure}

Now, we consider an optical emitter with an emission wavelength $\lambda_0=780$~nm, e.g., a quantum dot, a molecule, or a rubidium (Rb) atom, in the vicinity of the (dielectric) core-shell (Ag) nanoparticle.
The atomic dipole emits at a fixed frequency $\omega_0$ (wavelength $\lambda_0$) and it is not affected by the intensity of the excitation electric field, but only by the sphere, as depicted in Fig.~\ref{fig1}.

\begin{figure}[htbp]
\includegraphics[width=\columnwidth]{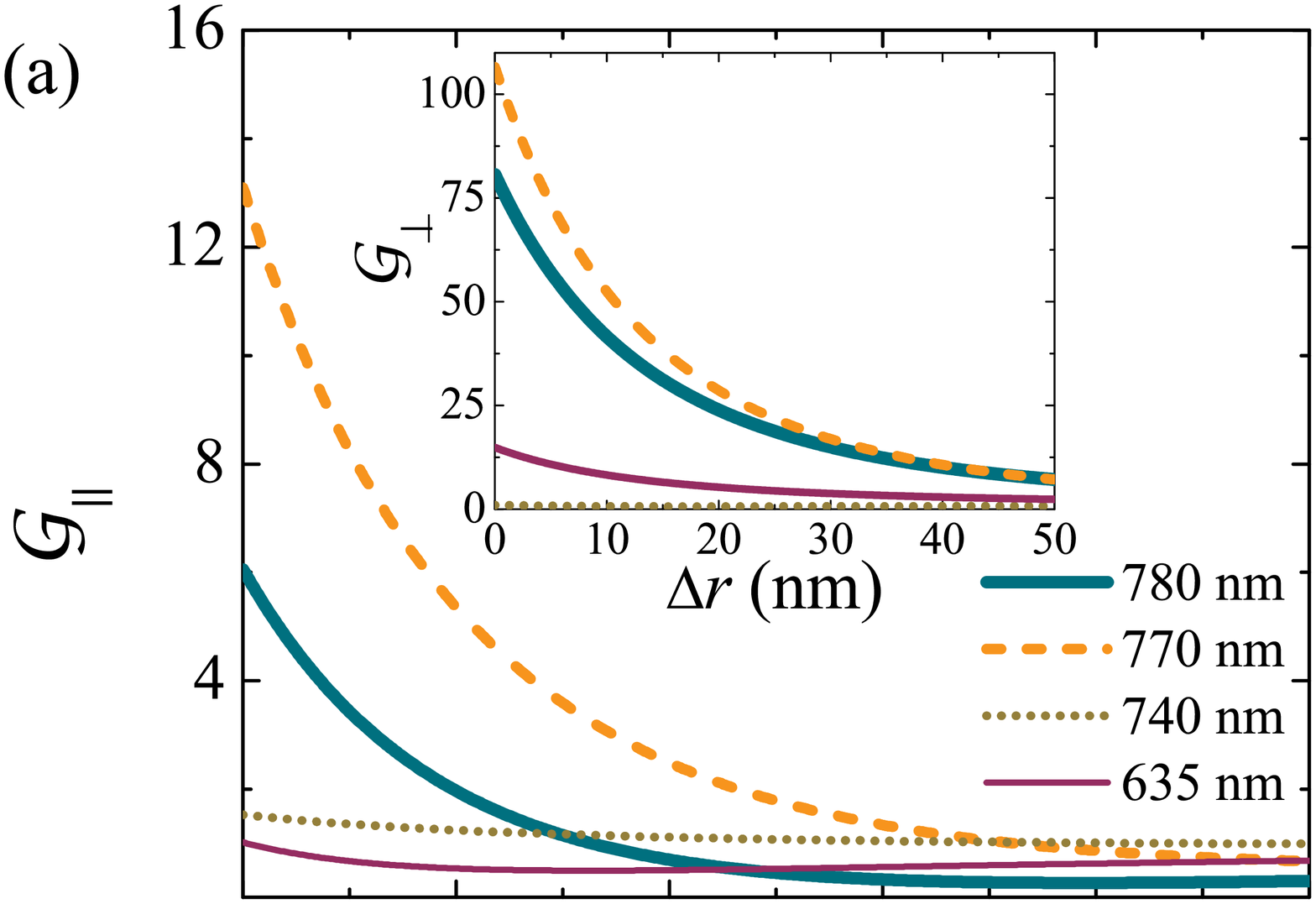}\vspace{-1.8cm}
\includegraphics[width=\columnwidth]{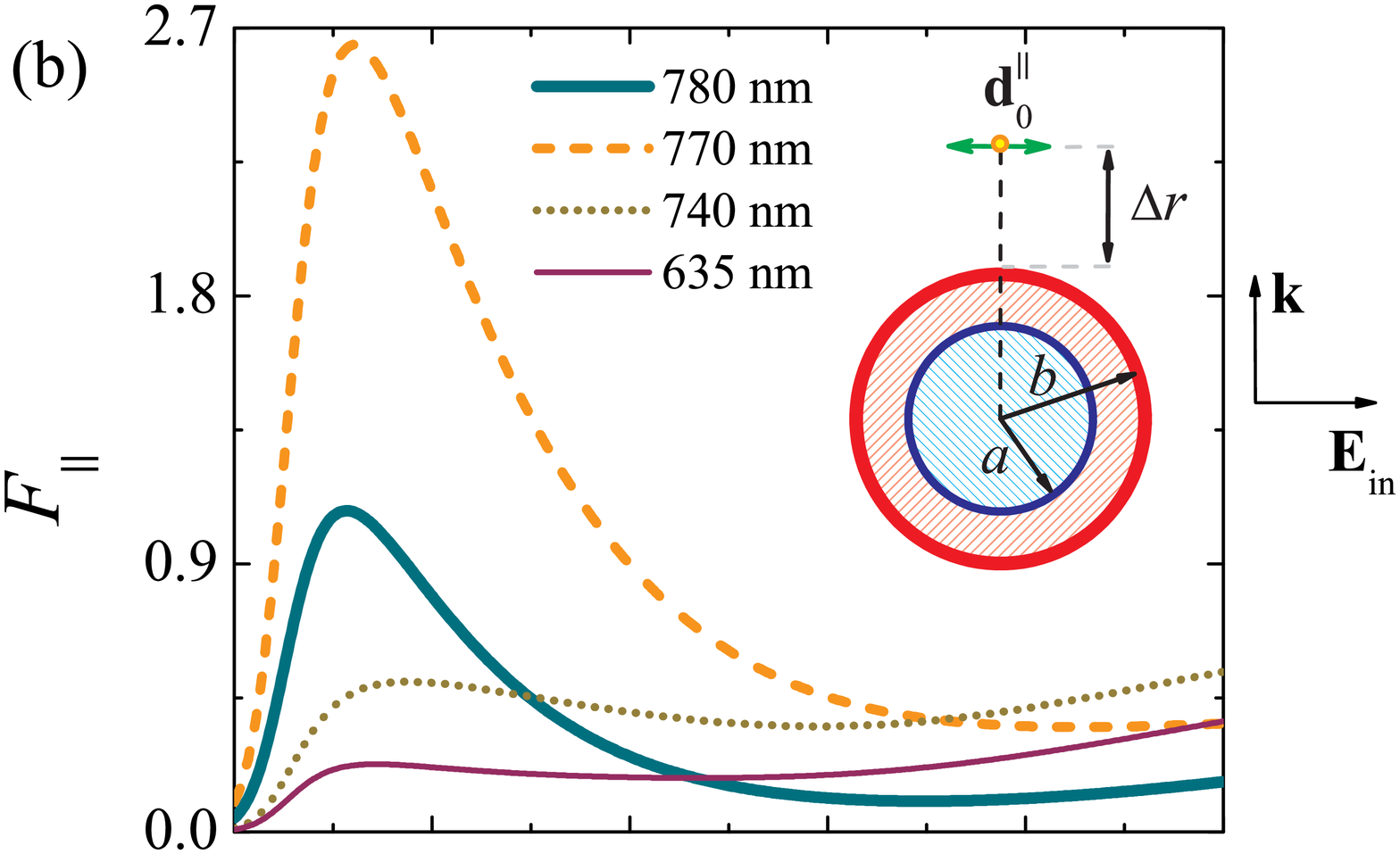}\vspace{-1.8cm}
\includegraphics[width=\columnwidth]{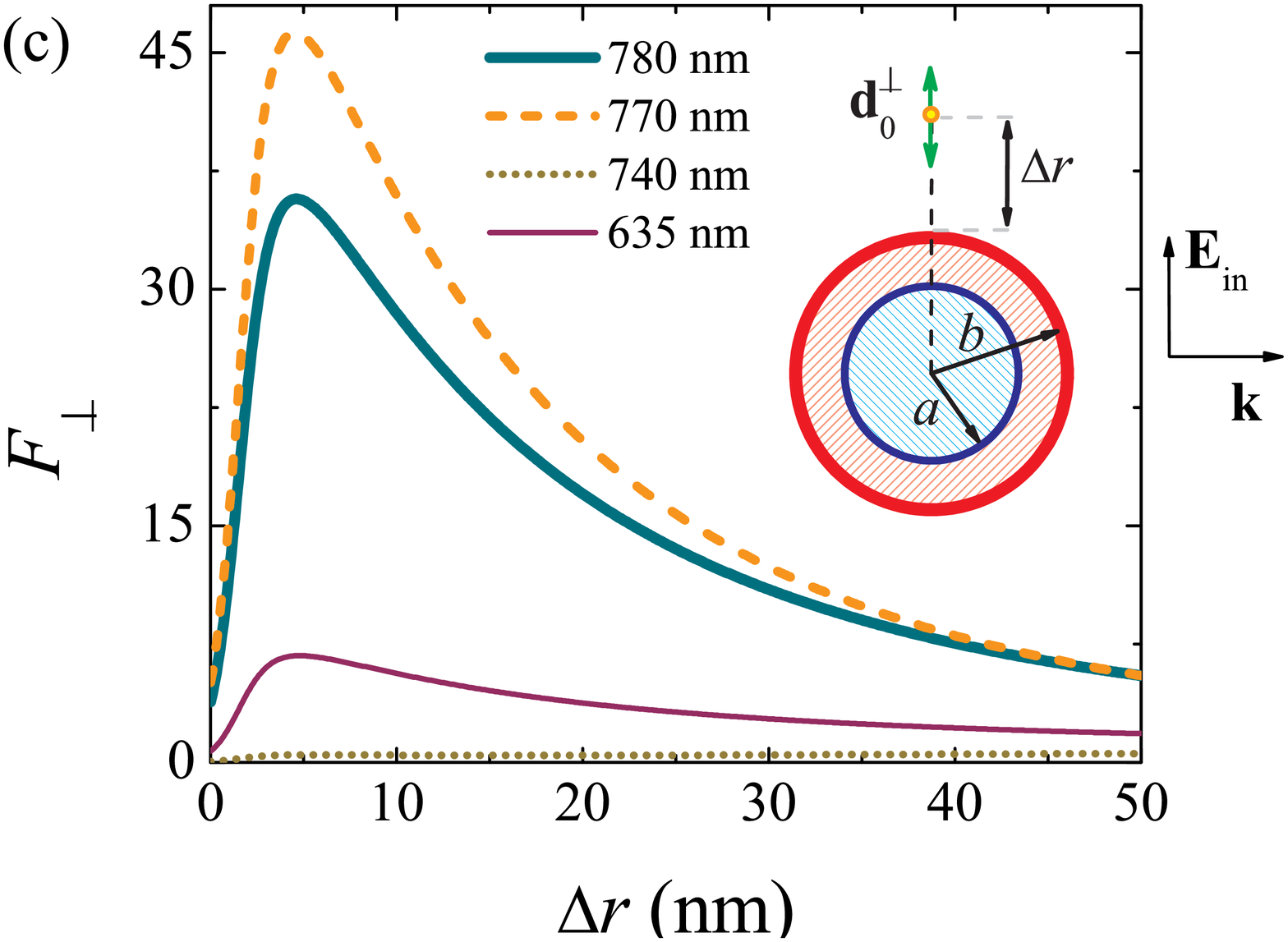}
\caption{The intensity and fluorescence enhancement factors for a quantum emitter with transition wavelength $\lambda_0=780$~nm as a function of the distance $\Delta r$ to a silver nanoshell and excitations wavelengths.
The core-shell nanoparticle consists of a dielectric core ($n_1=3.5$) with radius $a=50$~nm and a silver shell with radius $b=70$~nm.
The excitation wavelengths $\lambda$ are chosen from Fig.~\ref{fig2}, with $\mathbf{E}_{\rm in}(\lambda)$ being parallel to $\mathbf{d}_0^{\perp(||)}$: $780$~nm (emission wavelength), 770~nm (scattering resonance), 740 nm (cloaking or Fano dip in the scattering response), 635~nm (quadrupole resonance).
The plots show the intensity enhancement factors (a) $\mathcal{G}_{||}$ and $\mathcal{G}_{\perp}$ (inset), and the corresponding fluorescence enhancement factors (b) $F_{||}$ and (c) $F_{\perp}$.
The maximum fluorescence enhancement occurs for $\Delta r\approx 5$~nm.}\label{fig4}
\end{figure}

\begin{figure*}[htbp!]
{\includegraphics[width=\columnwidth]{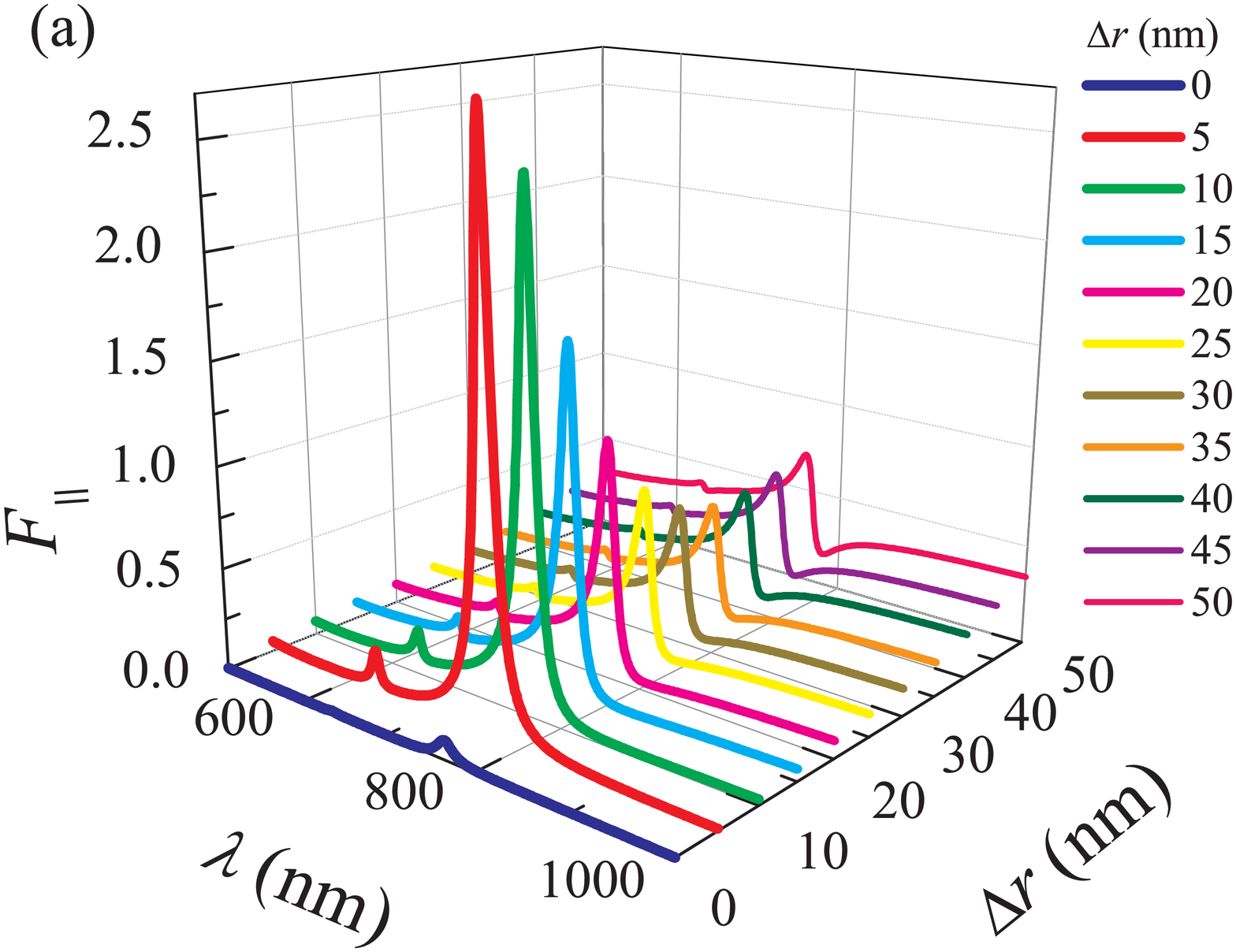}}
{\includegraphics[width=\columnwidth]{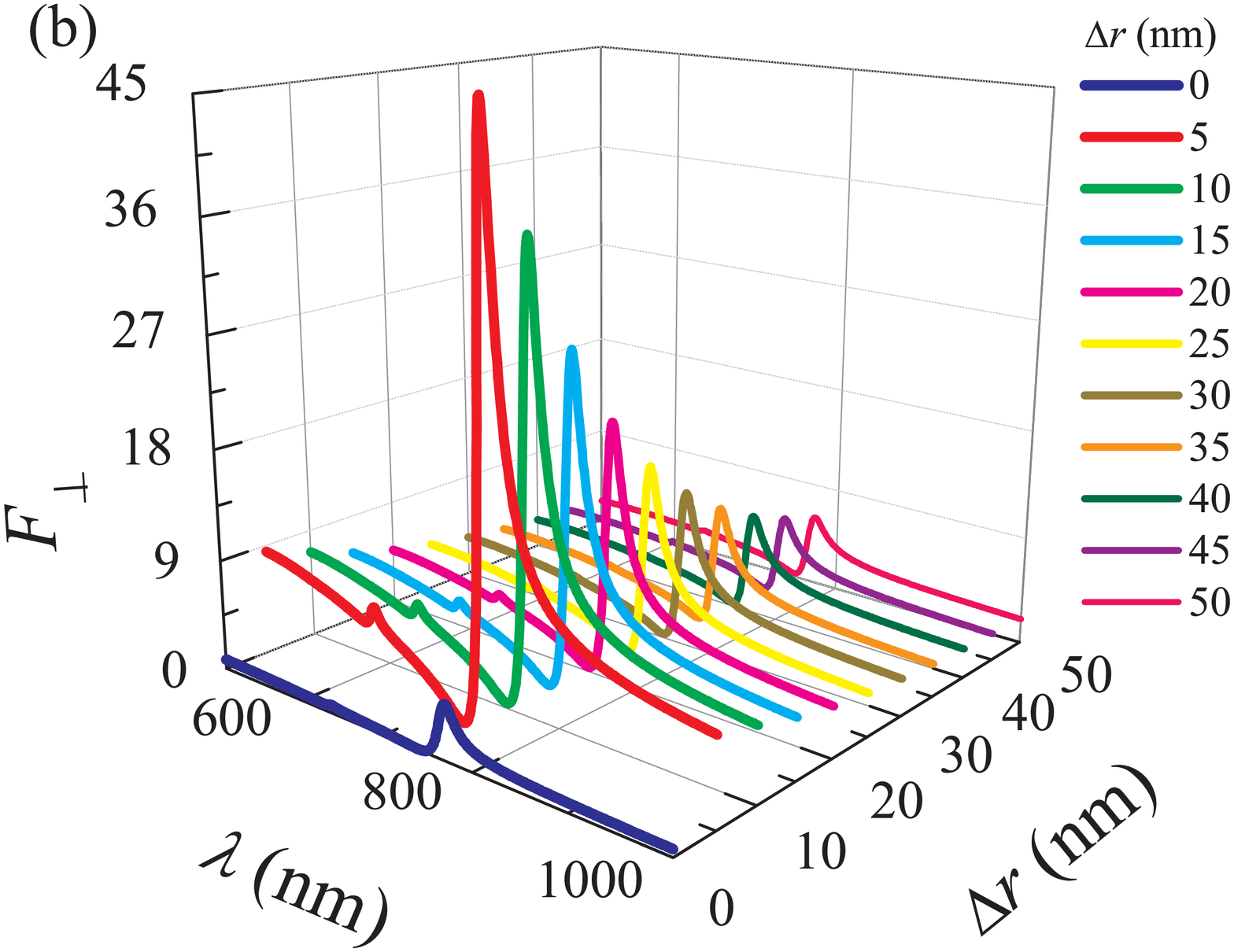}}
{\includegraphics[width=\columnwidth]{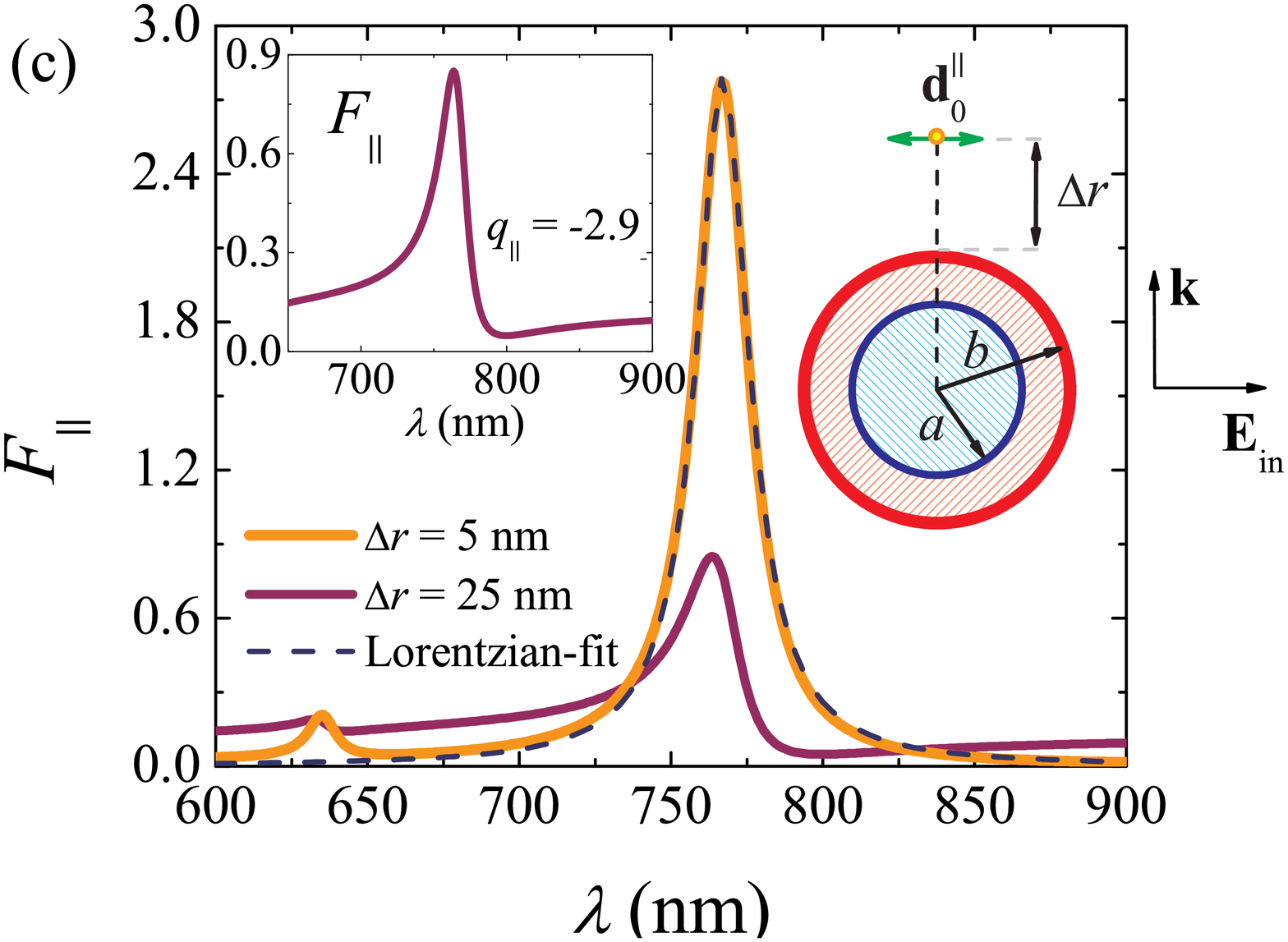}}
{\includegraphics[width=\columnwidth]{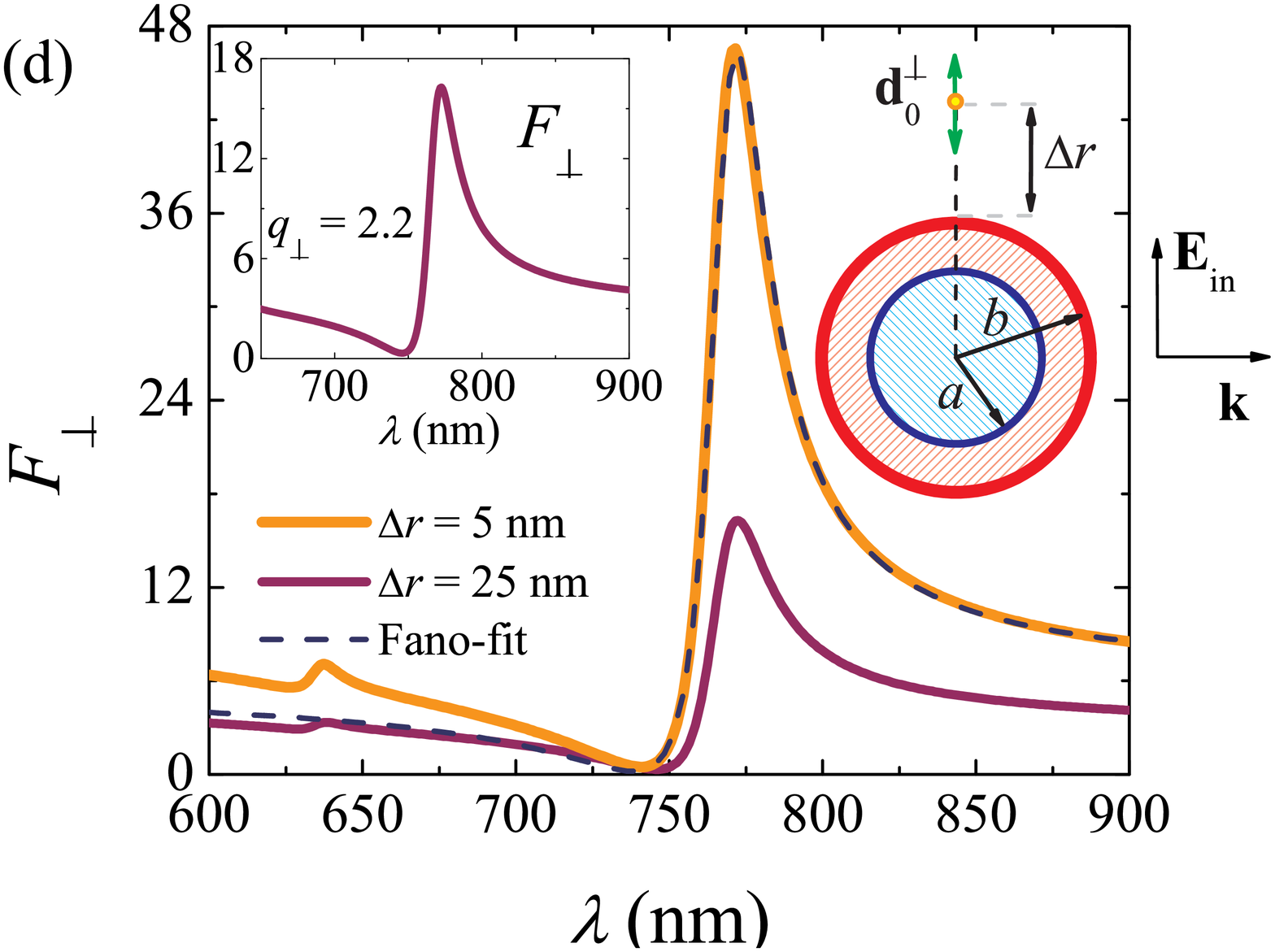}}
\caption{Fluorescence enhancement response associated with a quantum emitter (transition wavelength $\lambda_0=780$~nm) in the vicinity of a silver nanoshell.
The inner sphere has refractive index $n_1=3.5$ and radius $a=50$~nm, whereas the silver nanoshell has radius $b=70$~nm.
The plots show the fluorescence enhancement factor as a function of the excitation wavelength and the distance between the emitter and the spherical surface for a (a) tangentially and (b) normally oriented dipole, with $\mathbf{d}_0^{\perp(||)}$ being parallel to $\mathbf{E}_{\rm in}$.
From left to right, the peaks correspond to distances $\Delta r$ ranging from 0~nm (the thicker line) to $50$~nm (the thinner line).
(c) The plot shows the Lorentzian lineshape of $F_{||}$ for $\Delta r=5$~nm (maximum enhancement).
The inset shows a Fano lineshape for $\Delta r=25$~nm with negative asymmetry parameter $q_{||}=-2.9$.
(d) The plot shows the Fano lineshape of $F_{\perp}$ for $\Delta r=5$~nm (maximum enhancement).
The inset shows a Fano lineshape for $\Delta r=25$~nm with positive asymmetry parameter $q_{\perp}=2.2$.}\label{fig5}
\end{figure*}

Figures~\ref{fig3}(a) and \ref{fig3}(b) show the spontaneous-decay rates, calculated from the Lorenz-Mie theory, and the corresponding quantum efficiency of the system.
As can be seen in Fig.~\ref{fig3}(a), the tangential orientation of the atomic dipole moment is associated with a slow radiative decay rate, whereas a high radiative decay rate is obtained for the normal orientation of the atomic dipole moment.
Indeed, in the near field, the dipole moment with tangential orientation induces a dipole moment in the plasmonic nanoshell, but with opposite direction and almost the same amplitude, resulting in near cancellation of the effective electric field~\cite{Zadkov_PhysRevA85_2012}.
As a result, $\Gamma_{||}^{\rm rad}\ll\Gamma_{\perp}^{\rm rad}$ in the vicinity of the nanoshell.

Figure~\ref{fig3}(b) shows that when the atom lies on the nanoshell surface ($\Delta r=0$~nm), the nonradiative decay channel is dominant ($Q_{\perp}\approx 0$ and $Q_{||}\approx0$).
This is due to the near-field interaction of the atomic dipole moment with both bright ($\ell=1$) and dark $(\ell>1)$ electromagnetic modes, with the latter being the dominant interaction.
In a more complete and realistic picture, the atom sets up a polarization field due to its intrinsic polarizability, leading to a very strong interaction with the plasmonic surface via van der Waals forces, dispersion forces, and Casimir-Polder forces~\cite{Slama_NatPhys10_2014}.
Also, note that $Q_{\perp}>Q_{||}$ for our set of parameters, which means that a more efficient coupling between the dipole and the nanoshell is found for a normally oriented dipole.
Both quantum efficiencies, $Q_{\perp}$ and $Q_{||}$, are below unity due to ohmic losses in the metallic shell.
However, for large distances $r_0\gg b$, the inset in Fig.~\ref{fig3}(b) shows that both $Q_{\perp}$ and $Q_{||}\to1$. 

In Figs.~\ref{fig4}(a)-\ref{fig4}(c), we plot the intensity enhancement factor $\mathcal{G}$ and the corresponding fluorescence enhancement factor $F$ for some specific excitation wavelengths obtained from Fig.~\ref{fig2} as a function of $\Delta r$.
The wavelengths are 780~nm (transition wavelength of the optical emitter), 770~nm (dipolar scattering/absorption resonance in the core-shell sphere), 740~nm (scattering antiresonance) and 635~nm (quadrupolar scattering/absorption resonance).
The rigorous calculation usually considers that $\mathbf{d}_0$ is parallel to the local electric field, which takes into account both the incoming and scattered electric fields (for details, see Appendix~\ref{Intensity})~\cite{Zadkov_PhysRevA85_2012}. 
In the far-field region or in the absence of the sphere, $\mathbf{d}_0$ is parallel to the incoming electric field, $\mathbf{E}_{\rm in}(kz)=E_0e^{\imath k z}\hat{\mathbf{x}}$.
Indeed, for quantum emitters, the dipole orientation can be easily controlled by the polarization of the incoming electromagnetic wave~\cite{Slama_NatPhys10_2014}. 
To simplify our discussion,   we consider only the two basic dipole moment orientations, so that one can assume that the dipole emitter lies along the $x$-axis ($x>b$) for $\mathbf{d}_0^{\perp}$ and is perpendicular to the $z$-axis ($z>b$) for $\mathbf{d}_0^{||}$.

For all the excitation wavelengths in Figs.~\ref{fig4}(b) and \ref{fig4}(c), the maximum $F_{\perp(||)}$ occurs around $\Delta r\approx 5$~nm, despite the strong intensity enhancement factor $\mathcal{G}_{\perp(||)}$ for $\Delta r\to 0$, Fig.~\ref{fig4}(a).
This optimum fluorescence enhancement for $\Delta r\not=0$ is a consequence of optical absorption in the metallic nanoshell. 
Below $\Delta r\approx 5$~nm, nonradiative scattering channels dominate over the radiative one $(\ell=1)$, so that the atomic dipole strongly interacts with subradiant (dark) modes, which possess angular momentum with $\ell>1$.
Indeed, we have verified that the Lorenz-Mie series associated with the quantum efficiencies converges only for $\ell\gg1$ ($\ell_{\rm max}\approx 76$ for our set of parameters), despite the scattering cross section being well described by the dipole approximation ($\ell=1$), Fig.~\ref{fig2}. 
This agrees with previous theoretical and experimental results that show that the dipole approximation fails to describe the quenching of fluorescence for short distances~\cite{Novotny_PhysRevLett96_2006}. 
 
As expected from Fig.~\ref{fig3}(b), we have in Fig.~\ref{fig4} an efficient coupling for the normally oriented dipole in comparison with the tangentially oriented dipole, so that $F_{\perp}$ is one order of magnitude greater than $F_{||}$.
Due to the plasmonic Fano resonance, the fluorescence enhancement $F_{\perp}\approx48$ occurs even with a detuning of 10 nm from the atomic transition.
This result is encouraging from the experimental point of view since small variations in the resonance frequency of an ensemble of nanoshell structures are expected due to fabrication and material imperfections.  

\begin{figure}[htbp!]
\includegraphics[width=1.2\columnwidth]{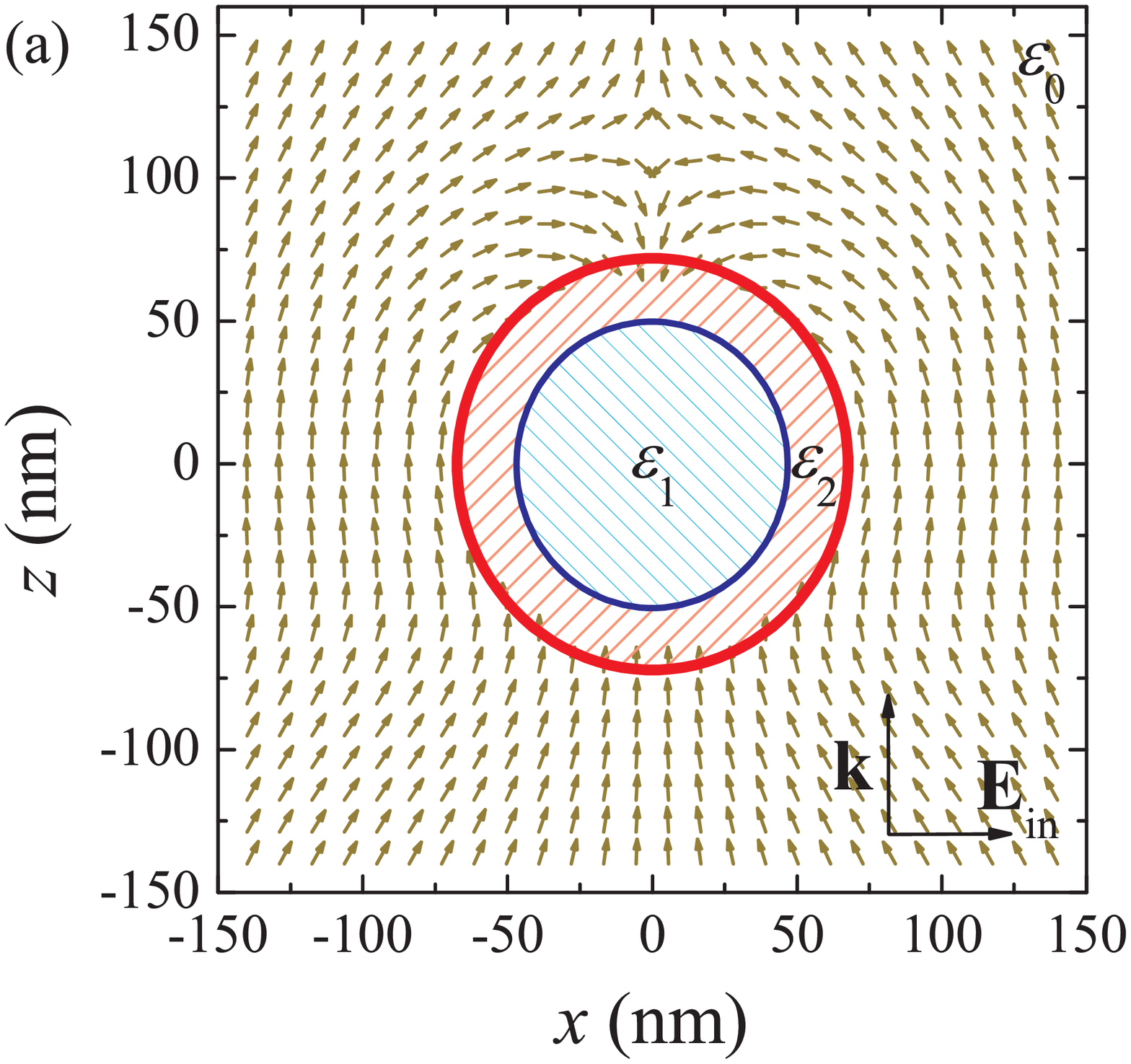}\hspace{-1cm}
\includegraphics[width=1.2\columnwidth]{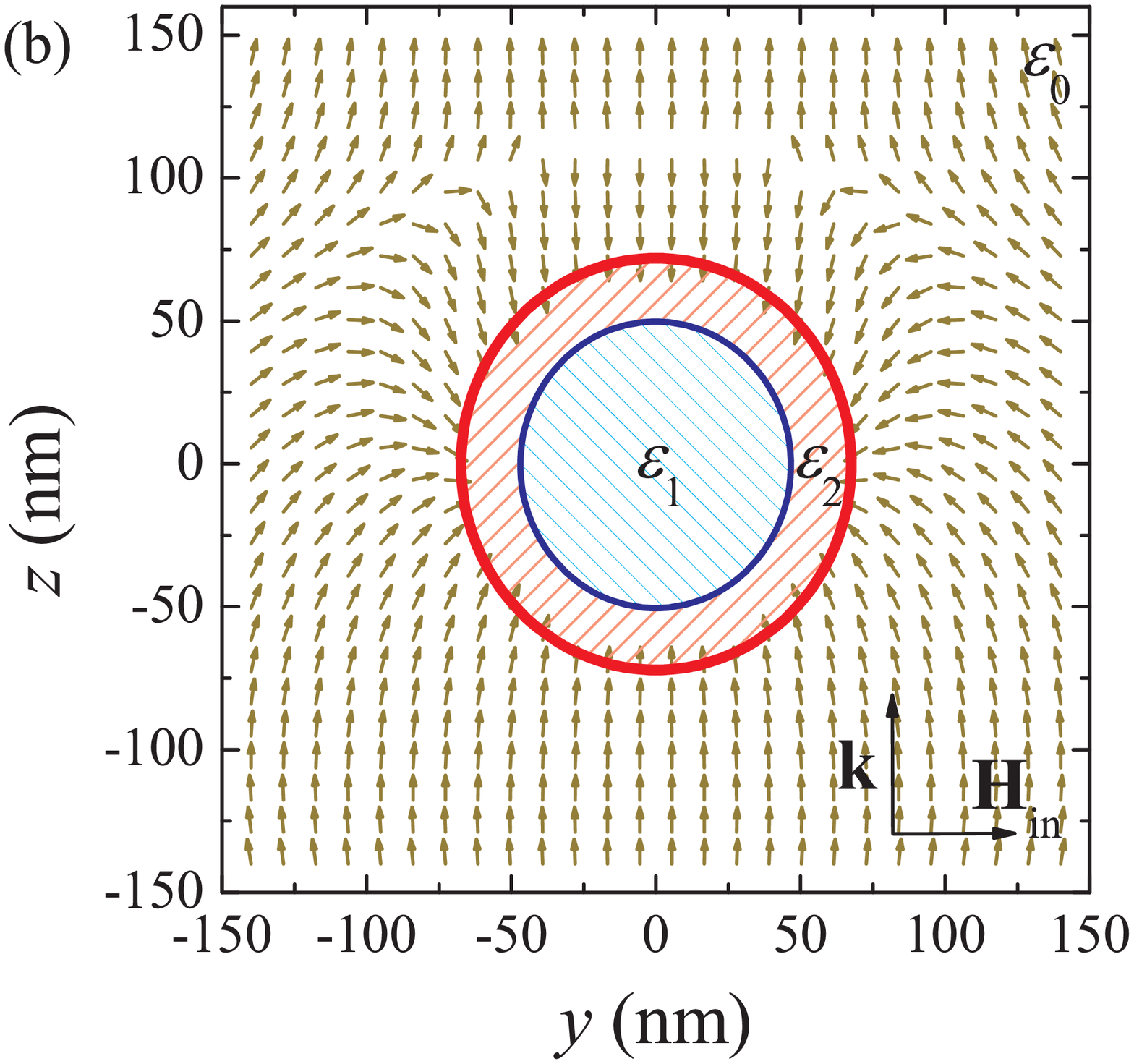}\hspace{-1cm}
\caption{Energy flow vector field (normalized Poynting vector) in the vicinity of a plasmonic Ag nanoshell interacting with an electromagnetic plane wave ($\lambda=770$~nm).
The dielectric core has refractive index $n_1=3.5$ and radius $a=50$~nm, whereas the Ag shell has radius $b=70$~nm.
(a) The $xz$ plane shows the presence of a saddle point in the energy flow in the $z$-axis around $z\approx 110$~nm ($\Delta r = 40$~nm).
(b) The $yz$ plane shows the singular point along the $y$ direction, $z\approx 110$~nm.}\label{fig6}
\end{figure}

In Figs.~\ref{fig5}(a)-\ref{fig5}(d), we plot the fluorescence enhancement factor as a function of the excitation wavelength.
Note that the $F_{||}$ profile changes continuously from a Lorentzian lineshape $(\Delta r=5$~nm) to a Fano lineshape as we increase $\Delta r$, whereas $F_{\perp}$ remains a Fano lineshape profile from the beginning to the end.
This change in $F_{||}$ behavior seems to occur around $\Delta r=20$~nm and $\Delta r=30$~nm, which is a local minimum region for the quantum efficiency $Q_{||}$ [see Fig.~\ref{fig3}(b)].
Furthermore, being a Fano lineshape described by $f_q(\epsilon) = (q+\epsilon)^2/(1+\epsilon^2)$, where $q$ is the Fano asymmetry parameter, the Fano profiles obtained in Figs.~\ref{fig5}(c) and \ref{fig5}(d) have asymmetry parameters with different sign, namely $q_{\perp}\approx 2.2$ and $q_{||}\approx -2.9$. 
These two plots are highlighted in the insets of Figs.~\ref{fig5}(c) and \ref{fig5}(d).

The sign difference between $q_{\perp}$ and $q_{||}$ and the transition between Lorentzian and Fano lineshapes appear in association with a singular point in the near-field energy flow induced by the Fano effect~\cite{Miroshnichenko_OptLett37_2012,Kivshar_JOpt15_2013}.
This connection between Fano resonances and singular points in the energy flow is pointed out in Ref.~\cite{Kivshar_JOpt15_2013}.
The plasmonic Fano resonance in a weakly absorbing subwavelength sphere is expected to be associated with optical vortices and saddle points~\cite{Alu_JOSAB24_2007} in the Poynting vector field around the particle in the vicinity of the resonance.
Indeed, as we show in Fig.~\ref{fig6} for $\lambda=770$~nm, there is a saddle point in the energy flow around the Ag nanoshell.
Since the incoming wave polarization is fixed along the $x$-axis, the fluorescence enhancement plotted in Figs.~\ref{fig5}(a) and \ref{fig5}(b) is probing the energy flow in the $x$- and $z$-axes, respectively.
In particular, the tangentially oriented dipole seems to be sensitive to saddle points in the energy flow, changing the response from a Lorentzian lineshape ($|q_{||}|\to\infty$) to a Fano profile ($|q_{||}|<\infty$) when passing through it.
A changing in the profile of the quantum efficiency $Q_{||}$ can also be observed around $\Delta r\approx 30$~nm in Fig.~\ref{fig3}(b), which is approximately the position in the $z$-axis where the saddle point is for the transition frequency.

To explain the difference between the two basic dipole orientations, we observe that one can rewrite the electric Mie coefficient $a_1(\omega)$ in the vicinity of a Fano resonance as~\cite{Tribelsky_PhysRevA93_2016}
\begin{align}
a_1(\omega)=\frac{\epsilon(\omega) + q}{\epsilon(\omega) + q - \imath\left[\epsilon(\omega)q-1\right]},\label{a1}
\end{align}
where $\epsilon(\omega)=\epsilon'(\omega)+\imath\epsilon''(\omega)$ and 
\begin{align}
q=\frac{\chi_1'(kb)}{\psi_1'(kb)},\label{q}
\end{align}
where $\psi_1(kb)=kbj_1(kb)$ and $\chi_1(kb)=-kby_1(kb)$ are the Riccati-Bessel and Riccati-Neumann functions, respectively, with $j_1$ and $y_1$ being the spherical Bessel and Neumann functions of first order.
If the sphere is lossless, one has $\epsilon''(\omega)=0$ and $|a_1|^2=(\epsilon+q)^2/[(1+q^2)(\epsilon^2+1)]$, i.e., $|a_1|^2$ is a normalized Fano lineshape and $q$ is the Fano asymmetry parameter of the scattering cross section curve.
The function $\epsilon(\omega)$ (not to be confused with the permittivity $\varepsilon$) can be obtained from Eq.~(\ref{an}) and is discussed in detail for homogeneous spheres in Ref.~\cite{Tribelsky_PhysRevA93_2016}.
In the vicinity of a Fano resonance, one can approximate this function by $\epsilon(\omega)\approx (\omega - \omega_{\rm res})/\Omega$, where $\Omega$ is associated with the curve linewidth.

\begin{figure}[htbp!]
\includegraphics[width=\columnwidth]{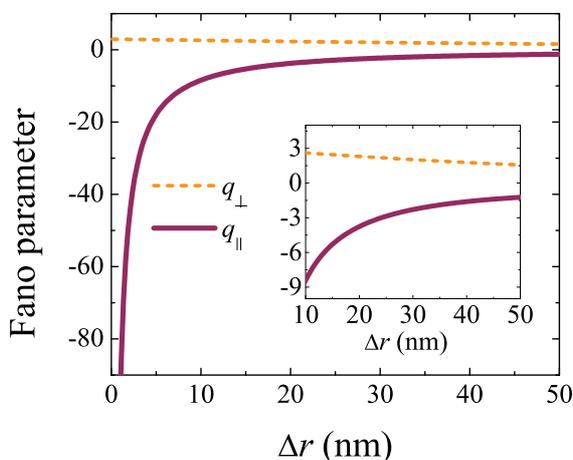}
\caption{The Fano asymmetry parameter of the fluorescence enhancement as a function of the distance between dipole and sphere.
The system and parameters are the same as in Fig.~\ref{fig5}.
The two curves provide the values of $q_{\perp}^{\lambda}=-q_{\perp}^{\omega}$ [Eq.~(\ref{q-perp})] and $q_{||}^{\lambda}=-q_{||}^{\omega}$ [Eq.~(\ref{q-para})] that fit the plots in Figs.~\ref{fig5}(a) and \ref{fig5}(b), respectively. 
The parameters used in Eq.~(\ref{eps-imag}) are $\sigma_{\rm sca}^{(\rm max)}\approx 5.4\pi b^2$, $\lambda_{\rm res}\approx 770$~nm, and $q=7.53$, leading to $\epsilon''\approx 0.83$.
To fit the Fano curves in Fig.~\ref{fig5}, we consider $\epsilon(\lambda)\approx(\lambda - 770)/10$ ($\lambda$ in nanometers) and the calculated asymmetry parameters.
The inset shows $q_{\perp(||)}$ in the range $10~{\rm nm}<\Delta r<50$~nm.
}\label{fig7}
\end{figure}

Substituting Eq.~(\ref{a1}) into Eqs.~(\ref{G-factor-perp}) and (\ref{G-factor-para}) for $\ell=1$, which is the dominant scattering channel for $kb<1$, we obtain after some algebra the Fano asymmetry parameters of the fluorescence enhancement curves:
\begin{align}
q_{\perp}&=\frac{1}{1+\epsilon''(\omega)}\left[\frac{qy_1(kr_0)-j_1(kr_0)}{qj_1(kr_0)+y_1(kr_0)}\right],\label{q-perp}\\
q_{||}&=-\frac{1}{1+\epsilon''(\omega)}\left[\frac{q \chi_1'(kr_0)+\psi_1'(kr_0)}{q\psi_1'(kr_0)-\chi_1'(kr_0)}\right].\label{q-para}
\end{align}
The function $\epsilon''(\omega)$ has a very complicated analytical expression, and it can be estimated from the maximum scattering cross section or the Fano resonance $[\epsilon'(\omega_{\rm res})=0]$: $\sigma_{\rm sca}^{(\rm max)} =6\pi(q^2 + \epsilon''^2)/[(k^2(1+q^2)(1+\epsilon''^2)]$.
Explicitly, we have the approximation
\begin{align}
\epsilon''(\omega_{\rm res}) \approx -1+\frac{q}{k}\sqrt{\frac{6\pi}{(1+q^2)\sigma_{\rm sca}^{\rm(max)}}}\label{eps-imag}
\end{align}

Equations~(\ref{q-perp}) and (\ref{q-para}) are the main result of this paper and describe the dependence of the Fano asymmetry parameters of the fluorescence enhancement on the distance between optical emitter and nanoshell. 
Since the analytical forms of $\mathcal{G}_{\perp(||)}(\omega)$ and $\Gamma_{\perp(||)}^{\rm rad}(\omega_0)/\Gamma_0$ are the same, Eqs.~(\ref{q-perp}) and (\ref{q-para}) can also be applied to describe the Fano effect on the Purcell factor of an optical emitter in the vicinity of a plasmonic nanoshell, where $\omega_0$ is a variable~\cite{Kivshar_SciRep5_2015}. 
Observe that these expressions are calculated in the frequency domain $\omega$, so that the Fano asymmetry parameters as a function of $\lambda$ have opposite sign: $q_{\perp(||)}^{\lambda} = -q_{\perp(||)}^{\omega}$.
Also, in the vicinity of a Fano resonance, we can write $\epsilon(\lambda)\approx (\lambda-\lambda_{\rm res})/\Lambda$, where $\Lambda$ is the linewidth.
In particular, from Eqs.~(\ref{q}) and (\ref{q-para}), we obtain that $r_0=b$ $(\Delta r=0)$ implies $|q_{||}|\to\infty$, i.e., a Lorentzian lineshape response.
This Lorentzian resonance continuously develops to a Fano resonance as a function of $\Delta r$.
This result is illustrated in Fig.~\ref{fig7}, where the plots of the  Fano asymmetry parameters show that $|q_{||}|<10$ for $\Delta r> 10$~nm.
The maximum degree of asymmetry, with $q_{\perp}^{\lambda}=1$ and $q_{||}^{\lambda}=-1$, is achieved for $\Delta r>50$~nm.

The strong symmetric Lorentzian peak at the vicinity of the nanoshell, for the tangential dipole configuration ($q_{||}\to\pm\infty$), means that only the narrow, localized plasmon mode is contributing to the fluorescence enhancement.
This occurs irrespective of the Fano resonance in the total scattering cross section.
Indeed,  with $q$ being the Fano asymmetry parameter of the cross sections,   one can easily show that $|q|\to\infty$ in $\sigma_{\rm sca}(\lambda)$ implies $|q_{||}|\to\infty$ in $F_{||}(\lambda)$, which is an expected result.
Conversely, for the dipole oriented normal to the spherical surface, one can obtain $|q_{\perp}|<\infty$ even for $|q|\to\infty$, which is usually the case for homogeneous nanoparticles with moderate permittivity.
This leads to a Fano lineshape in $F_{\perp}$ as a function of the excitation wavelength even for homogeneous metallic spheres, with Fano asymmetry parameter $q_{\perp}\approx [y_1(kr_0)/j_1(kr_0)]/(1+\epsilon'')$ (see, e.g., Figs.~9 and 10 of Ref.~\cite{Gaponenko_JPhysChem116_2012}).
In this case, the Fano resonance in $F_{\perp}(\lambda)$ is due to the interference between the angle-averaged scattered $[a_{\ell}(\omega)h_{\ell}^{(1)}(kr_0)]$ and incident [$j_{\ell}(kr_0)$] partial waves, where the latter is the nonresonant background.
If the spherical particle is a perfectly conducting (PC) homogeneous sphere of refractive index $n\gg1$, i.e., ${{a_{\ell}}(\omega)}_{n\gg1}\to a_{\ell}^{\rm PC}(\omega)\equiv\psi_{\ell}'(kb)/\xi_{\ell}'(kb)$ and ${b_{\ell}(\omega)}_{n\gg1}\to b_{\ell}^{\rm PC}(\omega)\equiv j_{\ell}(kb)/h_{\ell}^{(1)}(kb)$~\cite{Bohren_Book_1983,Tribelsky_PhysRevA93_2016}, one can achieve a Fano lineshape in $F_{\perp}(\lambda)$ and $F_{||}(\lambda)\approx 0$ in the near field [see Eqs.~(\ref{G-factor-perp}) and (\ref{G-factor-para})].

Here, the Lorentzian lineshape in $F_{||}(\lambda)$ in the near-field that changes into a Fano lineshape in the far-field is a consequence of the (dielectric) core-shell (metal) geometry. 
Physically, the atomic dipole moment $\mathbf{d}_{0}^{||}$ induces an oppositely directed dipole moment on the plasmonic nanoshell surface, with almost the same amplitude.
This interaction cancels out the broad dipole mode at the surrounding medium/plasmonic shell interface, but does not cancel out the narrow dipole mode at the plasmonic shell/dielectric core interface.
Explicitly, we can rewrite the electric Lorenz-Mie coefficient $a_{\ell}(\omega)$ as
\begin{align}
a_{\ell}(\omega) = a_{\ell}^{\rm PC}(\omega) - \frac{\left[\psi_{\ell}'(n_2kb)g_{\ell}(\omega) - \chi_{\ell}'(n_2kb)w_{\ell}(\omega)\right]}{n_2\xi_{\ell}'(kb)},\label{a-ell} 
\end{align}
where $g_{\ell}(\omega)$ and $w_{\ell}(\omega)$ are the Lorenz-Mie coefficients of the electromagnetic fields within the plasmonic shell as defined in Refs.~\cite{Bohren_Book_1983,Arruda_PhysRevA87_2013}.
The first term, $a_{\ell}^{\rm PC}(\omega)=\psi_{\ell}'(kb)/\xi_{\ell}'(kb)$, is related to the broad electric dipole mode $(\ell=1)$, while the second term accounts for the narrow electric dipole mode related to the plasmonic shell/core interface.
It is easily confirmed from Eqs.~(\ref{G-factor-perp}) and (\ref{G-factor-para}) that the term $a_1^{\rm PC}(\omega)$ in Eq.~(\ref{a-ell}) is canceled out for $r_0=b$ only in the intensity enhancement factor $\mathcal{G}_{||}(kr_0)$, leading to a Lorentzian lineshape response.
Also, the dipole moment of tangential orientation changes the phase of the narrow dipole mode in $\pi$.
As the distance between the dipole and the nanoshell becomes greater, the influence of the broad dipole mode in the fluorescence enhancement increases.
This influence is maximal beyond the near-field saddle point in the energy flow, as indicated in Figs.~\ref{fig5} and \ref{fig7}, and by the Poynting vector field in Fig.~\ref{fig6}.
From the experimental point of view, this result can be applied to control both the enhancement and quenching of the fluorescence response of quantum emitters in the vicinity of plasmonic nanoshells.

\section{Conclusion}
\label{Conclusion}

Based on the Lorenz-Mie theory, we have investigated the fluorescence enhancement of an optical emitter in the vicinity of a plasmonic silver nanoshell in the weak-coupling regime.
We have demonstrated that a Fano resonance in the total scattering cross section leads to a Fano lineshape response in the fluorescence enhancement as a function of the distance between dipole and sphere.
For an optical emitter with dipole moment oriented tangentially to the spherical surface, we have obtained a symmetric Lorentzian lineshape response in the near-field for the fluorescence enhancement.
In the far field, this  Lorentzian lineshape response changes into a Fano resonance, with Fano asymmetry parameter of opposite sign compared to the dipole moment oriented normally to the spherical surface.
This effect has been explained by the different role played by the induced electric dipole moment in the plasmonic nanoshell for both atomic dipole orientations.
We have shown that this change in the fluorescence enhancement can be calculated analytically and is also associated with an optical singular point in the energy flow in the vicinity of the plasmonic nanoshell.
These analytical results shed a light on a fundamental problem of Fano-like resonances in nanoplasmonics, and they may have interesting applications for fluorescence enhancement and/or quenching of optical dipole emitters near plasmonic nanoshells.

\section*{Acknowledgments}
%The authors thank John Weiner for the fruitful collaboration, discussions and suggestions to improve the manuscript. 
The authors acknowledge the Brazilian and German agencies for financial support.
T.J.A, R.B, and Ph.W.C hold Grants from S\~ao Paulo Research Foundation (FAPESP) (Grant Nos. 2015/21194-3, 2014/01491-0, and 2013/04162-5, respectively).
S.S is supported by the Fulbright-Cottrell Award.

\appendix

\section{Lorenz-Mie theory}
\label{Lorenz-Mie}

The incident and scattered electric fields provided by the Lorenz-Mie theory (plane waves) are~\cite{Bohren_Book_1983}:
\begin{align}
\mathbf{E}_{\rm in}(\mathbf{r})&=-\frac{1}{kr}\sum_{\ell=1}^{\infty}E_{\ell}\bigg\{\imath\cos\varphi\sin\theta j_{\ell}(kr)\ell(\ell+1)\pi_{\ell}\hat{\mathbf{r}}\nonumber\\
&-\cos\varphi\left[\pi_{\ell}\psi_{\ell}(kr)-\imath\tau_{\ell}\psi_{\ell}'(kr)\right]\hat{\boldsymbol{\theta}}\nonumber\\
&-\sin\varphi\left[\imath\pi_{\ell}\psi_{\ell}'(kr)-\tau_{\ell}\psi_{\ell}(kr)\right]\hat{\boldsymbol{\varphi}}\bigg\},\label{Ein}\\
\mathbf{E}_{\rm sca}(\mathbf{r})&=\frac{1}{kr}\sum_{\ell=1}^{\infty}E_{\ell}\bigg\{\imath\cos\varphi\sin\theta a_{\ell} h_{\ell}^{(1)}(kr)\ell(\ell+1)\pi_{\ell}\hat{\mathbf{r}}\nonumber\\
&-\cos\varphi\left[b_{\ell}\pi_{\ell}\xi_{\ell}(kr)-\imath a_{\ell}\tau_{\ell}\xi_{\ell}'(kr)\right]\hat{\boldsymbol{\theta}}\nonumber\\
&-\sin\varphi\left[\imath a_{\ell}\pi_{\ell}\xi_{\ell}'(kr)-b_{\ell}\tau_{\ell}\xi_{\ell}(kr)\right]\hat{\boldsymbol{\varphi}}\bigg\},\label{Esca}
\end{align}
where $E_{\ell} = \imath^{\ell} E_0(2\ell+1)/[\ell(\ell+1)]$, $\pi_{\ell}=P_{\ell}^1(\cos\theta)/\sin\theta$, $\tau_{\ell}={\rm d}P_{\ell}^1(\cos\theta)/{\rm d}\theta$, with $P_{\ell}^1$ being the associated Legendre function of first order.

The coefficients $a_{\ell}$ and $b_{\ell}$ are the electric (TM) and magnetic (TE) Lorenz-Mie coefficients, respectively, and are determined from boundary conditions.
For center-symmetric coated spheres interacting with plane waves, as depicted in Fig.~\ref{fig1}, these coefficients read~\cite{Bohren_Book_1983,Arruda_JOpt14_2012}:
\begin{align}
        a_{\ell} &=\frac{\widetilde{n}_2\psi_{\ell}'(kb)-\psi_{\ell}(kb)\mathcal{A}_{\ell}(n_2kb)}{\widetilde{n}_2\xi_{\ell}'(kb)-\xi_{\ell}(kb)\mathcal{A}_{\ell}(n_2kb)},\label{an}\\
        b_{\ell} &=\frac{\psi_{\ell}'(kb)-\widetilde{n}_2\psi_{\ell}(kb)\mathcal{B}_{\ell}(n_2kb)}{\xi_{\ell}'(kb)-\widetilde{n}_2\xi_{\ell}(kb)\mathcal{B}_{\ell}(n_2kb)},\label{bn}
\end{align}
with the auxiliary functions
\begin{align*}
\mathcal{A}_{\ell}(n_2kb)&=\frac{\psi_{\ell}'(n_2kb)-A_{\ell}\chi_{\ell}'(n_2kb)}{\psi_{\ell}(n_2kb)-A_{\ell}\chi_{\ell}(n_2kb)},\\
\mathcal{B}_{\ell}(n_2kb)&=\frac{\psi_{\ell}'(n_2kb)-B_{\ell}\chi_{\ell}'(n_2kb)}{\psi_{\ell}(n_2kb)-B_{\ell}\chi_{\ell}(n_2kb)},\\
        A_{\ell} &= \frac{\widetilde{n}_2\psi_{\ell}(n_2ka)\psi_{\ell}'(n_1ka)-\widetilde{n}_1\psi_{\ell}'(n_2ka)\psi_{\ell}(n_1ka)}{\widetilde{n}_2\chi_{\ell}(n_2ka)\psi_{\ell}'(n_1ka)-\widetilde{n}_1\chi_{\ell}'(n_2ka)\psi_{\ell}(n_1ka)}\
        ,\\
        B_{\ell}&=\frac{\widetilde{n}_2\psi_{\ell}'(n_2ka)\psi_{\ell}(n_1ka)-\widetilde{n}_1\psi_{\ell}(n_2ka)\psi_{\ell}'(n_1ka)}{\widetilde{n}_2\chi_{\ell}'(n_2ka)\psi_{\ell}(n_1ka)-\widetilde{n}_1\chi_{\ell}(n_2ka)\psi_{\ell}'(n_1ka)}\
        ,
\end{align*}
where the functions $\psi_{\ell}(z)=z j_{\ell}(z)$, $\chi_{\ell}(z)=-z y_{\ell}(z)$ and $\xi_{\ell}(z)=\psi_{\ell}(z)-\imath\chi_{\ell}(z)$ are the Riccati-Bessel, Riccati-Neumann and Riccati-Hankel functions, respectively, with $j_{\ell}$ and $y_{\ell}$ being the spherical Bessel and Neumann functions~\cite{Bohren_Book_1983}.
The relative refractive and impedance indices (in relation to the surrounding medium) are $n_p=\sqrt{\varepsilon_{p}\mu_{p}/(\varepsilon_0\mu_0)}$ and $\widetilde{n}_p = \sqrt{\varepsilon_{p}\mu_0/(\varepsilon_0\mu_{p})}$, with $p=\{1,2\}$~\cite{Arruda_JOpt14_2012}.
For nonmagnetic materials ($\mu_p=\mu_0$), one has $\widetilde{n}_p=n_p$~\cite{Arruda_JOSA27_1_2010}.
It is worth mentioning that these Lorenz-Mie coefficients for a single-layered center-symmetric sphere can be easily generalized to the multilayered case~\cite{Wang_RadSci26_1991}.

\section{Decay rates and Lorenz-Mie theory}
\label{Gamma-calculation}

Here we present an analytical method to calculate $\Gamma/\Gamma_0$ based on the Lorenz-Mie theory.
Instead of directly using the electric dipole fields $\mathbf{E}_{\rm dip}^{\mathbf{d}_0}(k_0r)$ (emitted) and $\mathbf{E}_{\rm sca}^{\mathbf{d}_0}(k_0r)$ (reflected)~\cite{Chew_JCPhys87_1987} and calculating the ratio of emitted power, $P/P_0$, we consider only the standard Lorenz-Mie solution for incident plane waves~\cite{Bohren_Book_1983} to calculate the projected electric LDOS and, hence, $\Gamma/\Gamma_0$.

Using the expansions of the electric fields provided in Appendix~\ref{Lorenz-Mie}, we define the projected radiative LDOS~\cite{Wildea_SurfSciRep70_2015,Gaponenko_JPhysChem116_2012}, $\rho_{\mathbf{d}_0}^{\rm rad}(r_0,\omega_0)/\rho_0$, as the total electric-field intensity in the near field due to the presence of the sphere normalized by the field intensity in the absence of the sphere, both of them projected along $\mathbf{d_0}$:
\begin{align}
\frac{\Gamma_{\mathbf{d}_0}^{\rm rad}(r_0,\omega_0)}{\Gamma_0}=\frac{\left\langle \left|\mathbf{d}_0\cdot\left[\mathbf{E}_{\rm in}(\mathbf{r}_0,\omega_0)+\mathbf{E}_{\rm sca}(\mathbf{r}_0,\omega_0)\right]\right|^2\right\rangle}{\left\langle\left|\mathbf{d}_0\cdot\mathbf{E}_{\rm in}(\mathbf{r}_0,\omega_0)\right|^2\right\rangle},\label{Gamma-method}
\end{align}
where $\langle \cdots \rangle = (1/{4\pi})\int_{-1}^{1}{\rm d}(\cos\theta)\int_0^{2\pi}{\rm d}\varphi (\cdots)$ is the angle average over 4$\pi$.
Observe that Eq.~(\ref{Gamma-method}) is equal to Eq.~(\ref{G-factor}) for $\omega=\omega_0$.
Here we are applying the computational procedure discussed in Refs.~\cite{Aguanno_PhysRevE69_2004,Gaponenko_JPhysChem116_2012}, in which the variation on the projected LDOS is equal to the variation on electromagnetic emitted power by a classical oscillator (normalized by the power emitted in vacuum) placed at $\mathbf{r}_0$ and for a specific orientation $\mathbf{d}_0$ with respect to the spherical surface.
Additional feedback terms between the dipole and the sphere (recurrent scattering) are expected to have negligible contribution to the electromagnetic fields~\cite{Kerker_AppOpt19_1980}.
It is worth emphasizing that Eq.~(\ref{Gamma-method}) provides the same result as Refs.~\cite{Chew_JCPhys87_1987,Ruppin_JCPhys76_1982}. 
In particular, the angle average in Eq.~(\ref{Gamma-method}) is a consequence of the definition of total radiated power.
Indeed, it is calculated by integrating the radial component of the Poynting vector associated with the electric dipole moment over a spherical surface with radius $r$ at the far-field ($r\to\infty$)~\cite{Chew_JCPhys87_1987}. 
 
Let us now consider two basic orientations for the electric dipole moment in spherical geometry:
\begin{align*}
\mathbf{d}_0^{\perp} = d_0\hat{\mathbf{r}}\ ,\quad \mathbf{d}_0^{||} = \frac{d_0}{\sqrt{2}}\left(\hat{\boldsymbol{\theta}}+\hat{\boldsymbol{\varphi}}\right),
\end{align*}
where $\mathbf{d}_0^{||}$ was chosen for convenience (indeed, it provides analytical solutions for the angular integrals).
Now, by substituting $\mathbf{E}_{\rm in}(k_0r_0)$ and $\mathbf{E}_{\rm sca}(k_0r_0)$ [Eqs.~(\ref{Ein}) and (\ref{Esca}), respectively] into Eq.~(\ref{Gamma-method}), we readily obtain the radiative decay rates,
\begin{align}
\frac{\Gamma_{\perp}^{\rm rad}(k_0r_0)}{\Gamma_{0}}&=\frac{3}{2}\sum_{\ell=1}^{\infty} \ell(\ell+1)(2\ell+1)\nonumber\\
&\times\left|\frac{j_{\ell}(k_0r_0)-a_{\ell}(\omega_0)h_{\ell}^{(1)}(k_0r_0)}{k_0r_0}\right|^2,\label{Gamma-perprad}\\
\frac{\Gamma_{||}^{\rm rad}(k_0r_0)}{\Gamma_{0}}&=\frac{3}{4}\sum_{\ell=1}^{\infty}(2\ell+1)\Bigg[\left|\frac{\psi_{\ell}'(k_0r_0)-a_{\ell}(\omega_0)\xi_{\ell}'(k_0r_0)}{k_0r_0}\right|^2\nonumber \\
&+\left|j_{\ell}(k_0r_0)-b_{\ell}(\omega_0)h_{\ell}^{(1)}(k_0r_0)\right|^2\Bigg].\label{Gamma-pararad}
\end{align}
where $\Gamma_{\perp}^{\rm rad}$ and $\Gamma_{||}^{\rm rad}$ refer to a dipole oscillating orthogonally ($\mathbf{d}_0^{\perp}$) or tangentially $(\mathbf{d}_0^{||})$ to the spherical surface, respectively.
To obtain Eqs.~(\ref{Gamma-perprad}) and (\ref{Gamma-pararad}), we have used the relations~\cite{Bohren_Book_1983}: $\int_{-1}^1{\rm d}(\cos\theta)(\pi_{\ell}\tau_{\ell'}+\tau_{\ell}\pi_{\ell'})=0$, $(2\ell+1)\int_{-1}^1{\rm d}(\cos\theta)(\pi_{\ell}\pi_{\ell'}+\tau_{\ell}\tau_{\ell'})=2\ell^2(\ell+1)^2\delta_{\ell\ell'}$ and $(2\ell+1)\int_{-1}^1{\rm d}(\cos\theta)\pi_{\ell}\pi_{\ell'}\sin^2\theta=2\ell(\ell+1)\delta_{\ell\ell'}$, with $\delta_{\ell\ell'}$ being the Kronecker delta.
The coefficients $a_{\ell}$ and $b_{\ell}$ are provided in Eqs.~(\ref{an}) and (\ref{bn}), respectively, in Appendix~\ref{Lorenz-Mie}.

The total decay rate, $\Gamma=\Gamma^{\rm rad}+\Gamma^{\rm nrad}$, can be calculated from the definition in Eq.~(\ref{equiv1}), by using the Green's tensor associated with the electric dipole response.
However, following Refs.~\cite{Chew_JCPhys87_1987,Dujardin_OptExp16_2008}, we can readily derive the total decay rate from the equations above by noting that $|a_{\ell}|^2={\rm Re}(a_{\ell})$ and $|b_{\ell}|^2={\rm Re}(b_{\ell})$ when the sphere is lossless, a consequence of the optical theorem~\cite{Bohren_Book_1983} [see Eqs.(\ref{Qext})-(\ref{Qabs})].
By this simple observation, we can heuristically calculate the total decay rate from the radiative decay rate. 
Expanding the squared terms in Eqs.~(\ref{Gamma-perprad}) and (\ref{Gamma-pararad}) and replacing $|a_{\ell}|^2$ with ${\rm Re}(a_{\ell})$ and $|b_{\ell}|^2$ with ${\rm Re}(b_{\ell})$, we obtain the total decay rates
\begin{align}
\frac{\Gamma_{\perp}(k_0r_0)}{\Gamma_{0}}&=1-\frac{3}{2}\sum_{\ell=1}^{\infty}\ell(\ell+1)(2\ell+1)\nonumber\\
&\times{\rm Re}\left\{a_{\ell}(\omega_0)\left[\frac{h_{\ell}^{(1)}(k_0r_0)}{k_0r_0}\right]^2\right\},\label{Gamma-perp}\\
\frac{\Gamma_{||}(k_0r_0)}{\Gamma_{0}}&=1-\frac{3}{4}\sum_{\ell=1}^{\infty}(2\ell+1){\rm Re}\Bigg\{a_{\ell}(\omega_0)\left[\frac{\xi_{\ell}'(k_0r_0)}{k_0r_0}\right]^2 \nonumber\\
&+ b_{\ell}(\omega_0) h_{\ell}^{(1)}(k_0r_0)^2\Bigg\},\label{Gamma-para}
\end{align}
where we have used the relation $\sum_{\ell=1}^{\infty}\ell(\ell+1)(2\ell+1)j_{\ell}^2(z)=2z^2/3=\sum_{\ell=1}^{\infty}(2\ell+1)[\psi_{\ell}^2(z)+\psi_{\ell}'^2(z)]$.
Once again, assuming the dipole has no defined orientation in space, one has from Eqs.~(\ref{Gamma-perp}) and (\ref{Gamma-para}) the mean spatial ratio $\Gamma = (\Gamma_{\perp} + 2\Gamma_{||})/3$.

In addition, by subtracting Eqs.~(\ref{Gamma-perprad}) and (\ref{Gamma-pararad}) from Eqs.~(\ref{Gamma-perp}) and (\ref{Gamma-para}), respectively, we obtain the nonradiative decay rates
\begin{align}
\frac{\Gamma_{\perp}^{\rm nrad}(k_0r_0)}{\Gamma_{0}}&=\frac{3}{2}\sum_{\ell=1}^{\infty}
\ell(\ell+1)(2\ell+1)\left|\frac{h_{\ell}^{(1)}(k_0r_0)}{k_0r_0}\right|^2\nonumber\\
&\times{\rm Re}\left[a_{\ell}(\omega_0)-|a_{\ell}(\omega_0)|^2\right],\label{Gamma-perpn}\\
\frac{\Gamma_{||}^{\rm nrad}(k_0r_0)}{\Gamma_{0}}&=\frac{3}{4}\sum_{\ell=1}^{\infty}(2\ell+1)\nonumber\\
&\times{\rm Re}\Bigg\{\left|\frac{\xi_{\ell}'(k_0r_0)}{k_0r_0}\right|^2\left[a_{\ell}(\omega_0)-|a_{\ell}(\omega_0)|^2\right]\nonumber\\
&+\left|h_{\ell}^{(1)}(k_0r_0)\right|^2\left[b_{\ell}(\omega_0)-|b_{\ell}(\omega_0)|^2\right]\Bigg\}.\label{Gamma-paran}
\end{align}
Although we have been discussing electric dipole radiation in the vicinity of a sphere, analogous expressions can be readily obtained for a magnetic dipole by interchanging $a_{\ell}$ with $b_{\ell}$~\cite{Chew_JCPhys87_1987}.

\section{Intensity enhancement factor}
\label{Intensity}

We consider the incoming electromagnetic wave polarized along the $x$-axis, {i.e.} $\mathbf{E}_{\rm in}(\omega)$ is parallel to $\hat{\mathbf{x}}=\sin\theta\cos\varphi\hat{\mathbf{r}}+\cos\theta\cos\varphi\hat{\boldsymbol{\theta}}-\sin\varphi\hat{\boldsymbol{\varphi}}$.
The atomic dipole moment $\mathbf{d}_0$ is directed along the direction of the local electric field, making an angle $\varsigma$ with $\hat{\mathbf{r}}$~\cite{Zadkov_PhysRevA85_2012}.
Hence, the radial and tangential dipole moments are $\mathbf{d}_0^{\perp}=d_0\hat{\mathbf{r}}$ and $\mathbf{d}_0^{||}=d_0(\hat{\boldsymbol{\theta}}+\hat{\boldsymbol{\varphi}})/\sqrt{2}$, respectively.
Following the definition in Eq.~(\ref{G-factor}) and using the results of Sec.~\ref{Gamma-calculation}, for the two possible orientations of the electric dipole in relation to the spherical surface, we have:
\begin{align}
\mathcal{G}_{\perp}(kr_0) &= \frac{3}{2}\sum_{\ell=1}^{\infty}\ell(\ell+1)(2\ell+1)\left|\frac{j_{\ell}(kr_0)-a_{\ell}(\omega)h_{\ell}^{(1)}(kr_0)}{kr_0}\right|^2,\label{G-factor-perp}\\
\mathcal{G}_{||}(kr_0)&=
\frac{3}{4}\sum_{\ell=1}^{\infty}(2\ell+1)\Bigg[\left|\frac{\psi_{\ell}'(kr_0)-a_{\ell}(\omega)\xi_{\ell}'(kr_0)}{kr_0}\right|^2\nonumber\\
&+\left|j_{\ell}(kr_0)-b_{\ell}(\omega)h_{\ell}^{(1)}(kr_0)\right|^2\Bigg].\label{G-factor-para}
\end{align}

The intensity enhancement factor for a dipole with an arbitrary orientation is $\mathcal{G}=(\mathcal{G}_{\perp}+2\mathcal{G}_{||})/3$, which agrees with the result of Ref.~\cite{Gaponenko_JPhysChem116_2012}.
The precise intensity enhancement factor for an atomic dipole located in an arbitrary position $\mathbf{r}_0$ is $\mathcal{G}(\mathbf{r}_0)=\cos^2\varsigma\ \mathcal{G}_{\perp}(r_0)+\sin^2\varsigma\ \mathcal{G}_{||}(r_0)$, where $\cos^2\varsigma=|E_{r}(\mathbf{r}_0)|^2/|\mathbf{E}(\mathbf{r}_0)|^2$ and $\sin^2\varsigma=1-|E_r(\mathbf{r}_0)|^2/|\mathbf{E}(\mathbf{r}_0)|^2$, with $|\mathbf{E}|^2=|E_r|^2+|E_{\theta}|^2+|E_{\varphi}|^2$ being the local electric field intensity~\cite{Zadkov_PhysRevA85_2012}.
If one assumes that $\mathbf{d}_0$ is parallel to $\hat{\mathbf{x}}$, one has $\mathcal{G}(\mathbf{r}_0)=\mathcal{G}_{\perp}(kx)$ for the radial atomic dipole and $\mathcal{G}(\mathbf{r}_0)=\mathcal{G}_{||}(kz)$ for the tangential atomic dipole.

\end{document}